\documentclass[twocolappendix]{emulateapj}

\newcommand{\kms}{\rm ~km~s^{-1}}

\newcommand{\hmpc}{~h^{3}~{\rm Mpc}^{-3}}

\slugcomment{\bf Last updated:\today}

\usepackage{color}
\usepackage{multirow}
\usepackage{float}

\begin{document}

\title{CATALOGS OF COMPACT GROUPS OF GALAXIES FROM THE ENHANCED SDSS DR12}

\author{Jubee Sohn$^{1,2}$, %\footnote{email: jsohn@cfa.harvard.edu}, 
        Margaret J. Geller$^{1}$, Ho Seong Hwang$^{3}$, H. Jabran Zahid$^{1}$, Myung Gyoon Lee$^{2}$ }

\affil{$^{1}$ Smithsonian Astrophysical Observatory, 60 Garden Street, Cambridge, MA 02138, USA}
\affil{$^{2}$ Astronomy Program, Department of Physics and Astronomy, Seoul National University, Gwanak-gu, Seoul 08826, Korea}
\affil{$^{3}$ School of Physics, Korea Institute for Advanced Study, 85 Hoegiro, Dongdaemun-Gu, Seoul 02455, Korea}

%=============================================================
\begin{abstract}
We apply a friends-of-friends algorithm 
 to an enhanced SDSS DR12 spectroscopic catalog including redshift from literature
 to construct a catalog of $1588~N\ge3$ compact groups of galaxies
 containing 5179 member galaxies and covering the redshift range $0.01 < z < 0.19$. 
This catalog contains 18 times as many systems 
 and reaches 3 times the depth of similar catalog of \citet{Bar96}. 
We construct catalogs from both magnitude-limited and volume-limited galaxy samples. 
Like \citet{Bar96} we omit the frequently applied isolation criterion
 in the compact group selection algorithm. 
Thus the groups selected by fixed projected spatial and 
 rest frame line-of-sight velocity separation produce 
 a catalog of groups with a redshift independent median size. 
In contrast with previous catalogs, 
 the enhanced SDSS DR12 catalog (including galaxies with $r < 14.5$) 
 includes many systems with $z\lesssim 0.05$. 
The volume-limited samples are unique to this study. 
The compact group candidates in these samples 
 have a median stellar mass independent of redshift. 
Groups with velocity dispersion $\lesssim 100 \kms$ show 
 abundant evidence for ongoing dynamical interactions among the members.  
The number density of the volume-limited catalogs agrees with previous catalogs 
 at the lowest redshifts but decreases as the redshift increases.
The SDSS fiber placement constraints 
 limit the catalog completeness. 
In spite of this issue the volume-limited catalogs
 provide a promising basis for detailed spatially resolved probes 
 of the impact of galaxy-galaxy interactions within 
 similar dense systems over a broad redshift range.
\end{abstract}
%=============================================================
\section{INTRODUCTION}

Compact groups of galaxies are the densest known systems
 typically containing three to ten galaxies 
 within only a few tens of kiloparsecs. 
Since the discovery of Stephan's Quintet \citep{Ste77}, 
 numerous studies have identified 
 small aggregations of galaxies. 
\citet{Ros77} and \citet{Hic82} first identified 
 well-defined samples of compact groups 
 in the local universe.
Since then many investigators have constructed 
 compact group catalogs based on much more extensive galaxy surveys
 \citep{Pra94, Bar96, All00, Foc02, Iov02, Iov03, Lee04, deC05, McC09, Dia12}.

Because the separation of galaxies within compact groups is comparable with their sizes,
 these dense systems have been laboratories 
 for the study of the impact of galaxy-galaxy interactions 
 using spatially resolved spectroscopy \citep{Rub91, Alf15}
 and multi-wavelength observations
 (e.g. \citealp{Bit10, Bit11, Bit14, Soh13, Wal10, Wal12, Wal16, Alat15, Fed15, Zuc16} and references therein). 
These groups are also associated with 
 extended X-ray emission \citep{Pon96, Fus13, Des14}. 
The low HI content of compact groups is probably a consequence of
 continuous tidal stripping or heating by gravitational interactions 
 among member galaxies \citep{Ver01, Mar12}. 
Low redshift compact groups are intriguing targets 
 for integral field unit (IFU) observations
 that examine internal kinematics along with, for example, 
 spatially resolved strong-line star formation and metallicity diagnostics
 that probe the timescale of the gravitational interactions \citep{Vog13, Vog15, Alf15}. 

The high frequency of obvious tidal interactions 
 among compact group members suggests that 
 their lifetime must be short \citep{Bar85, Dia94}. 
In principle, the galaxies should merge within a few Gyr timescale. 
This time scale is comparable to the group crossing time. 
Nonetheless, compact groups are abundant 
 in the nearby universe with space densities ranging from 
 $10^{-4} \hmpc$ to $10^{-6} \hmpc$ 
 \citep{MdO91, Bar96, Lee04, Men11, Pom12, Soh15}. 

The mere existence of compact groups 
 at the current epoch remains a puzzle. 
Some previous studies propose that compact groups may persist for a much longer
 than the apparent interactions would suggest \citep{Gov91, Dia94, Ath97}. 
On the other hand, \citet{Dia94} suggest that 
 the groups are replenished from the surrounding environment as galaxies merge. 
However, evaluation of the environments of compact groups 
 in existing catalogs paints a confusing picture. 
In some catalogs 50-76\% are in surroundings \citep{Ram94, Men11} dense enough for replenishment; 
 in the 2MASS compact group catalog only 27\% \citep{Dia15} are 
 in dense surroundings making the replenishment model untenable. 
 
Existing catalogs of compact groups contain some observational biases 
 which may limit their usefulness for understanding the existence and evolution of these systems. 
The selection 
 may introduce an artificial dependence of apparent group properties on redshift and 
 may bias the environments of the selected systems. 
\citet{Bar96} first derive
 a catalog of compact groups from a complete redshift survey. 
They use a straightforward friends-of-friends (FoF) algorithm 
 to select group members separated by a fixed projected separation and rest frame velocity difference. 
As a result of the selection, 
 their catalog includes both nearby groups and groups in dense surroundings. 
They emphasize that previous criteria designed to select 
 isolated systems actually introduce a bias against the inclusion of the nearest compact group candidates. 
 
Here we extend the approach of \citet{Bar96} to the 
 sample of Sloan Digital Sky Survey (SDSS) Data Release 12 (DR12) at $r<17.77$
 (corresponding to the main galaxy sample of SDSS DR7). 
To capture nearby systems we enhance the SDSS redshift survey 
 by including galaxies with $r < 14.5$. 
We use a similar FoF algorithm to select 
 a sample of 1588 compact group candidates
 from the magnitude-limited catalog. 
This sample of compact group candidates is large enough 
 to enable the construction of volume-limited subsamples. 
These samples are important for understanding observational biases 
 including the SDSS fiber-positioning constraint. 
These samples are also important for exploring the properties of the groups 
 that are selected to be similar throughout the sample redshift range.

Section 2 describes the data 
 we use including the galaxy sample with $r < 14.5$.
We explain the FoF algorithm in \S 3. 
We also discuss the additional criteria applied by previous studies and 
 lay out the ways that these criteria may affect the resulting catalog. 
We describe the resulting magnitude- and volume-limited catalogs and
 compare our catalogs with a previous large catalog \citep{McC09} and 
 explore the salient differences.
In \S4, we compare the properties of compact group candidates in the catalogs we derive 
 with the properties of groups in previous samples. 
One important difference is that 
 groups in previous catalogs have sizes that tend to increase with redshift
 which is not seen in this study. 
In \S 5 we discuss the selection issues which may lead to this behavior. 
These selection issues also tend to eliminate nearby dense systems 
 that are prime candidates for detailed multi-wavelength studies.  
In \S 6 we use the volume-limited catalogs 
 to highlight the impact of the SDSS fiber-positioning constraints 
 on the selection of compact group candidates from the redshift survey. 
We include an appendix that discusses systems removed by 
 the additional selection on luminosity contrast and surface brightness. 
We conclude in \S 7. 
Throughout, we adopt the flat $\Lambda$CDM cosmological parameters of 
 $H_{0} = 100~h ~\rm km ~\rm s^{-1} ~\rm Mpc^{-1}$, 
 $\Omega_{m}=0.27$, and $\Omega_{\Lambda}=0.73$.

%=============================================================
\section{THE DATA: SDSS DR12}

\begin{figure}
\centering
\includegraphics[width=8.5cm,angle=0]{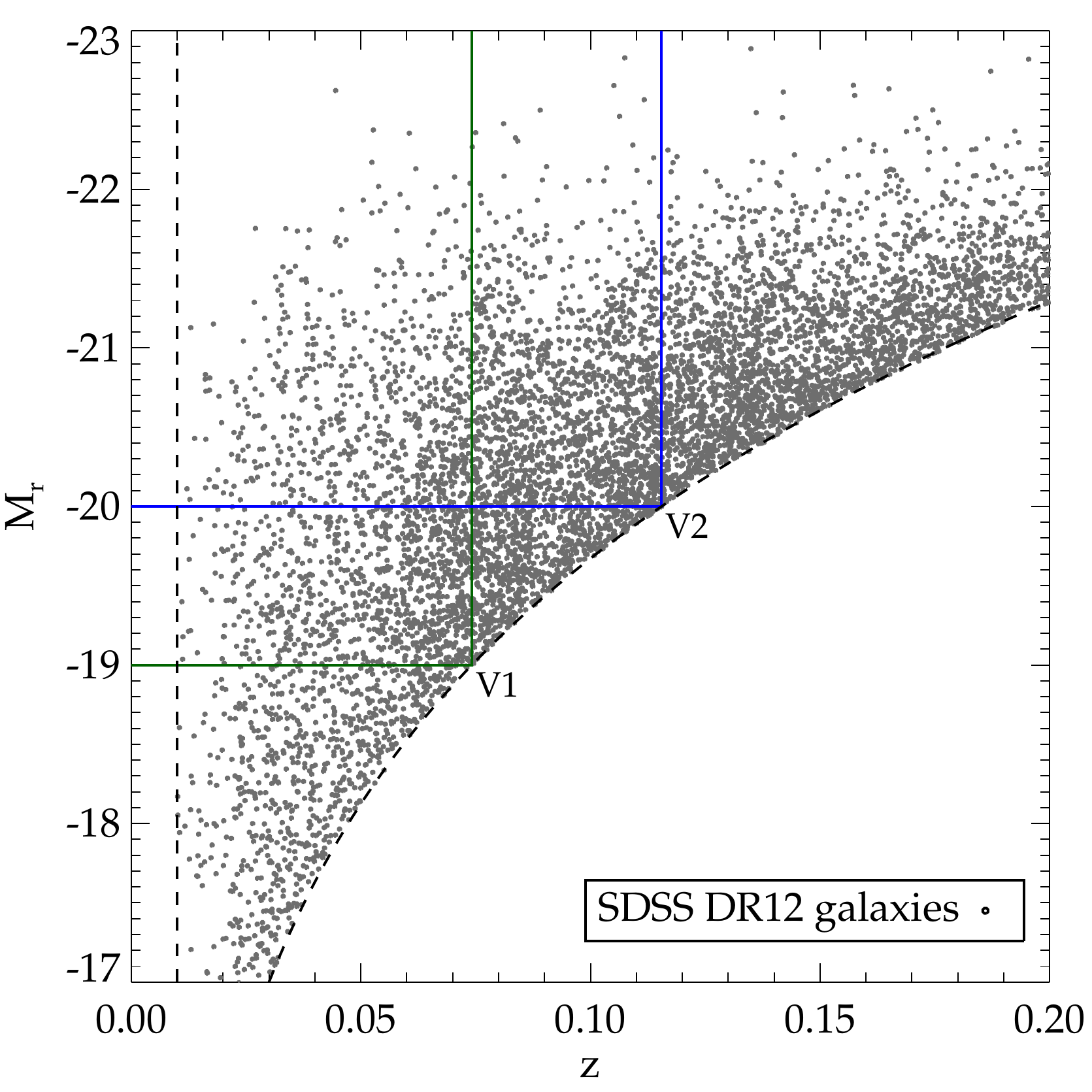}
\caption{
Absolute k-corrected $r-$band magnitudes of the SDSS DR12 galaxies
 as a function of redshift 
 (we display only 1\% of the data for clarity). 
Solid lines define two volume-limited samples: 
 V1 with $0.01 \leq z \leq 0.0754$ and $M_{r} \geq -19$ mag and 
 V2 with $0.01 \leq z \leq 0.1154$ and $M_{r} \geq -20$ mag. }
\label{volume}
\end{figure}

We derive a sample of compact groups from 
 the spectroscopic sample of SDSS DR12 \citep{Ala15} galaxies at $r<17.77$.
The DR12 includes redshifts for more than 2.4 million galaxies. 
The SDSS galaxy sample is incomplete for galaxies with $r < 14.5$
 because of the saturation \citep{Par09}. 
Fiber placement constraints also limit 
 the completeness of catalogs of compact groups derived from the SDSS.  

To reduce the incompleteness of the SDSS, 
 we supplement the catalog with redshifts from the literature
 (see \citealt{Hwa10} for details).
We also add redshifts 
 from recent FAST observations 
 at Fred Lawrence Whipple Observatory \citep{Soh15}. 
These additional data (FAST + literature) include 
 7796 redshifts for galaxies with $r < 14.5$ 
 and 22507 for galaxies with $14.5 \leq r < 17.77$.
The resulting sample contains 2,782,483 redshifts. 

Figure \ref{volume} shows
 the $K_{z=0.1}$ corrected, evolution-corrected 
 absolute $r-$band magnitudes as a function of redshift. 
We use the $K-$correct software (ver 4.2) of \citet{Bla07} for K-correction, 
 shifted to $z=0.1$, and applied the evolution correction given by \citet{Teg04},
 $E(z) = 1.6(z-0.1)$ \citep{Hwa12}. 
The dashed line indicates the magnitude limit of the sample.
We further restrict the catalog to galaxies 
 in the redshift range $0.01 < z < 0.20$. 
These limits remove the Virgo cluster and 
 they avoid use of the low density survey at the high redshift end of the SDSS.
This restricted redshift sample includes 654,066 galaxies.

We adopt the morphology for compact group galaxies 
 from the Korea Institute for Advanced Study (KIAS) 
 DR7 value-added catalog (VAGC) \citep{Choi10}. 
The KIAS DR7 VAGC lists the morphology of galaxies 
 based on the $u-r$ color, the $g-i$ color gradient, 
 and on the $i$ band concentration index \citep{Par05}. 
For the small fraction of galaxies not included in the KIAS DR7 VAGC, 
 we visually inspect the galaxies 
 and divide them into early and late types
 based on the SDSS images. 

We measure stellar masses using the LePHARE\footnote{http://www.cfht.hawaii.edu/$\sim$arnouts/LEPHARE/lephare.html} 
 spectral energy distribution (SED) fitting code \citep{Arn99, Ilb06}. 
The mass-to-light ratio is calculated 
 by fitting synthetic SEDs to the observed photometry.
We adopt the photometric parameters of compact group galaxies 
 from the SDSS pipeline \citep{Sto02}.
Synthetic SEDs are generated using the stellar population synthesis models of \citet{Bru03}. 
%(Bruzual & Charlot, MRNAS, 2003, 344, 1000). 
We vary the star formation history, age, extinction and metallicity of the stellar population. 
The star formation histories are exponentially 
 declining ($\propto e^{-t/\tau}$) with $e$-folding times ($\tau$) ranging 
 between 0.1 and 30 Gyr. 
The stellar population ages range between 0 and 13 Gyr. 
The \citet{Cal00} %(Calzetti et al. 2000, ApJ, 533, 682) 
 extinction law is adopted and $E(B-V)$ ranges from $0 - 0.6$. 
The models have two metallicities and 
 we use the \citet{Cha03} %(Chabrier, 2003, PASP, 115, 763} 
 initial mass function to calculate stellar mass.  

%=============================================================
\section{\label{selection}COMPACT GROUP SELECTION}

\subsection{The {\it friends-of-friends} algorithm}\label{fof}

We use the FoF algorithm of \citet{Bar96}
 to identify compact groups in the spectroscopic sample.
For each galaxy, the FoF finds neighboring galaxies  
 within a fixed projected physical spatial distance ($\Delta D$) and 
 rest frame line-of-sight velocity linking length ($\Delta V$). 
The linking lengths are redshift independent.   
We bundle linked galaxies 
 into a single galaxy system
 (e.g. \citealp{Tur76, Huc82, Bar96, Tag10, Rob11, Tem14}). 
The resultant compact group candidates 
 we identify have a projected physical size that is essentially redshift independent
 (see Section \ref{prop}). 

We test various linking lengths for the FoF. 
We start from 
 a projected linking length of $\Delta D \leq 50~h^{-1}$ kpc and 
 a radial linking length of $|\Delta V| = 1000 \kms$ 
 following \citet{Bar96} who identified compact groups 
 from the CfA2 and SSRS redshift surveys.
\citet{Bar96} used these linking lengths
 because they approximately matched the median separation between Hickson compact group galaxies. 
Their compact groups thus have physical properties 
 similar to the Hickson compact groups \citep{Bar96, Wal16}. 

The linking lengths of $\Delta D \leq 50~h^{-1}$ kpc and $|\Delta V| = 1000 \kms$
 we choose identify 42 of the 57 Hickson compact groups.
The groups we miss would require a larger linking length of
 $\Delta D \leq 125~h^{-1}$ kpc. 
In general, the projected linking length of $\Delta D = 50~h^{-1}$ kpc and 
 the rest frame line-of-sight velocity of $|\Delta V| \leq 1000 \kms$ \citep{Bar96}
 actually identify compact groups with physical sizes and galaxy number densities similar
 to the Hickson compact groups (see \S \ref{prop}). 
\citet{Woo10} demonstrate that 
 these linking lengths minimize  
 interlopers with discordant redshifts 
 while recovering systems similar to the original Hickson compact groups.  
These linking lengths also have the advantage that 
 they are often used to identify close pairs \citep{Bar00, Lin04}.
Thus, our catalog can be combined with catalogs of close pairs in the literature and 
 future analyses of these compact groups can be compared to previous work on pairs. 

%%=================================
%%%%   Table 1
%%=================================
\begin{turnpage}
\begin{table*}[h!]
  \begin{center}
    \caption{Catalog of MLCGs}
    \label{grtab}
    \begin{tabular}{cccccccccccc}
    \hline
    \multirow{2}{*}{ID} & R.A. & Decl & \multirow{2}{*}{$N_{mem}$} & 
    \multirow{2}{*}{z\tablenote{The error is the $1\sigma$ deviation derived from 1000-times bootstrap resamplings.}} & 
    $R_{gr}^{\rm a}$ & $\log \rho^{\rm a}$ & $\sigma_{CG}^{\rm a}$ & 
    \multirow{2}{*}{$N_{C}$\tablenote{Number of neighboring galaxies in a comoving cylinder of $\Delta R < 700~h^{-1}$ kpc and rest frame $|\Delta V| < 1000 \kms$.}} & 
    \multirow{2}{*}{Subset\tablenote{Designations for groups that contain sub-groups with a tighter radial linking length. `S' indicates a group containing a sub-group; `N' designates a group with no tighter subgroup. }} & 
    \multirow{2}{*}{V1CG subsample} & \multirow{2}{*}{V2CG subsample} \\
     & (J2000) & (J2000) & & & ($h^{-1}$ kpc) & ($h^{3}$ Mpc$^{-3}$) & ($\kms$) & & & & \\
    \hline 
    MLCG0001 & 251.559814 &  31.722052 &  3 & $0.0534 \pm 0.0005$ & $ 30.1 \pm  4.3$ & $4.42 \pm 0.29$ & $ 287 \pm  70$ &   1 & N &      -- &      --  \\
    MLCG0002 & 140.182175 &  33.686241 &  4 & $0.0229 \pm 0.0003$ & $ 36.8 \pm  2.9$ & $4.28 \pm 0.13$ & $ 170 \pm  31$ &   7 & S &      -- &      --  \\
    MLCG0003 & 146.524597 &  34.623241 &  3 & $0.1318 \pm 0.0010$ & $ 20.7 \pm  1.4$ & $4.91 \pm 0.10$ & $ 628 \pm 164$ &   0 & N &      -- &      --  \\
    MLCG0004 & 154.741531 &  37.298065 &  3 & $0.0480 \pm 0.0003$ & $ 32.9 \pm  5.5$ & $4.30 \pm 0.42$ & $ 210 \pm  37$ &   1 & S & V1CG003 &      --  \\
    MLCG0005 & 158.222275 &  12.086633 &  3 & $0.0330 \pm 0.0004$ & $ 12.3 \pm  3.2$ & $5.58 \pm 0.72$ & $ 242 \pm  71$ &   4 & S & V1CG004 &      --  \\
    \hline
    \end{tabular}
  \end{center}
\tablecomments{
The full table is available in the online journal. A portion is shown here for guidance regarding its form and content.}  
\end{table*}
\end{turnpage}
 
We also apply the FoF algorithm 
 with an even tighter rest frame line-of-sight linking length $|\Delta V| \leq 500 \kms$
 and indicate these groups in Table \ref{grtab}.
Previous studies of tight galaxy pairs showed that 
 pairs with this tighter separation 
 are more likely to be bound \citep{Bar00, Pat00, Haw03, deP07}.
The subset of compact groups 
 with this tighter radial linking length
 can be used for comparison with these previous galaxy pair samples.

Hickson and others (e.g. \citealp{Iov03, Lee04, deC05, McC09, Dia12})
 apply additional criteria to define compact groups. 
We apply the population and compactness criteria applied by \citet{McC09}. 
These criteria are a modified version of the original \citet{Hic82} approach.
\begin{itemize}
\item{The population limit requires $\geq 2$ additional members 
 within $\Delta r < 3~{\rm mag}$ of the brightest group member. 
 Here $r$ is the SDSS extinction- and $k-$corrected $r-$band model magnitude. 
 This criterion eliminates groups that contain 
 one dominant galaxy surrounded by much fainter satellite galaxies.}
\item{The compactness criterion, $\mu_{r} < 26$ mag arcsec$^{-2}$, 
 ($\mu_{r}$ is the $r-$band surface brightness averaged over the group radius) 
 excludes groups containing only low luminosity, 
 low surface brightness galaxies.}
\end{itemize}
In the Appendix, %\ref{app} 
 we briefly discuss the typical groups 
 that we eliminate with these additional criteria and comment on possible avenues 
 that could be explored further by including these systems. 
We do not apply an isolation criterion because, 
 as we discuss below, 
 this criterion artificially selects against nearby groups.
The isolation criterion applied by \citet{Hic82} and others requires that
 there be no other galaxies within an annulus of $R_{gr} < R_{nogal} < 3 R_{gr}$. 
Here $R_{gr}$ is the radius of the smallest circle encompassing all group members and 
 $R_{nogal}$ is the distance between the nearest non-member galaxy with 
 $\Delta r < 3~{\rm mag}$ and the group center. 

We identify groups consisting of at least three galaxies.
\citet{Hic82} originally defined compact groups
 with at least four member galaxies, 
 and some previous compact group surveys use his definition
 \citep{McC09, Men11, Dia12}.
However, subsequent spectroscopic observations show that
 many of the compact group candidates selected 
 photometrically contain only three member galaxies with accordant redshifts
 \citep{Hic92, Pom12, Soh15}.

\subsection{Catalogs of Compact Groups from a Complete Redshift Survey\label{cgsam}}

Most previous compact group catalogs have been extracted from magnitude-limited surveys
 (e.g. \citealp{Bar96, Iov03, McC09}). 
We also construct a sample from the SDSS DR12 magnitude-limited redshift survey 
 to obtain the largest possible sample of candidate compact groups (MLCG hereafter). 
In addition, we define two volume-limited samples of compact groups 
 to explore the selection biases inherent in the magnitude-limited catalog. 
The two volume-limited samples include (Figure \ref{volume}): 
 galaxies with $M_{r} < -19.0$ and $0.01 < z < 0.0741$ (V1),  
 galaxies with $M_{r} < -20.0$ and $0.01 < z < 0.1154$ (V2).
To construct volume-limited compact group catalogs, 
 we apply the FoF algorithm to the two volume-limited samples independently. 
Table \ref{sample} lists the number of groups in each catalog and 
 specifies the limiting survey parameters. 
 
%=================================
%%%   Table 2
%=================================

\begin{table*}
  \begin{center}
    \caption{The Compact Group Samples}
    \label{sample}
    \begin{tabular}{lcccccc}
    \hline
    Sample & Magnitude Limit & z range & $N_{\rm sample}$\tablenote{The number of galaxies in $N\ge4$ and $N=3$ compact groups, and chance alignments. } & $N\geq3$ CGs & $N\ge4$ CGs & $N=3$ CGs \\
    \hline
    MLCGs & $17.77~(m_{r})$ & [0.01, 0.20]   & 654066 & 1588 & 312 & 1276 \\
    V1CGs & $-19.0~(M_{r})$ & [0.01, 0.0741] & 149573 &  675 & 122 &  553 \\
    V2CGs & $-20.0~(M_{r})$ & [0.01, 0.1154] & 210834 &  298 &  36 &  261 \\
    \hline
    \end{tabular}
  \end{center}
\end{table*}    

We compare the physical properties of compact group candidates in the MLCG
 with previous catalogs that are 
 also selected from magnitude-limited samples 
 (e.g. \citealp{McC09, Soh15}).
We examine some of the selection biases in the MLCG 
 by comparing it with the two volume-limited subsets of the catalog, V1CG and V2CG 
 (Section \ref{vlsam}).

Fiber-positioning constraints in the SDSS introduce a systematic undersampling of regions 
 that are dense on the sky \citep{Str02, Par09, She16}. 
This undersampling leads to an incomplete catalog of compact group candidates 
 just as it leads to an incomplete sample of close pairs. 
\citet{She16} considered the impact of the SDSS DR6 incompleteness 
 on samples of close pairs with separations $\leq 100~h_{70}^{-1}$ kpc. 
They conclude that the fraction of missing pairs 
 increases steeply with redshift for $z > 0.09$.
Our volume-limited compact group candidate samples (Section \ref{vol}) 
 provide a measure of the bias 
 introduced by the SDSS incompleteness. 
In spite of the SDSS incompleteness, 
 the MLCG serves as a finding list, albeit incomplete, of candidate compact systems.

\subsubsection{The MLCG}

\begin{figure}
\centering
\includegraphics[width=8.5cm,angle=0]{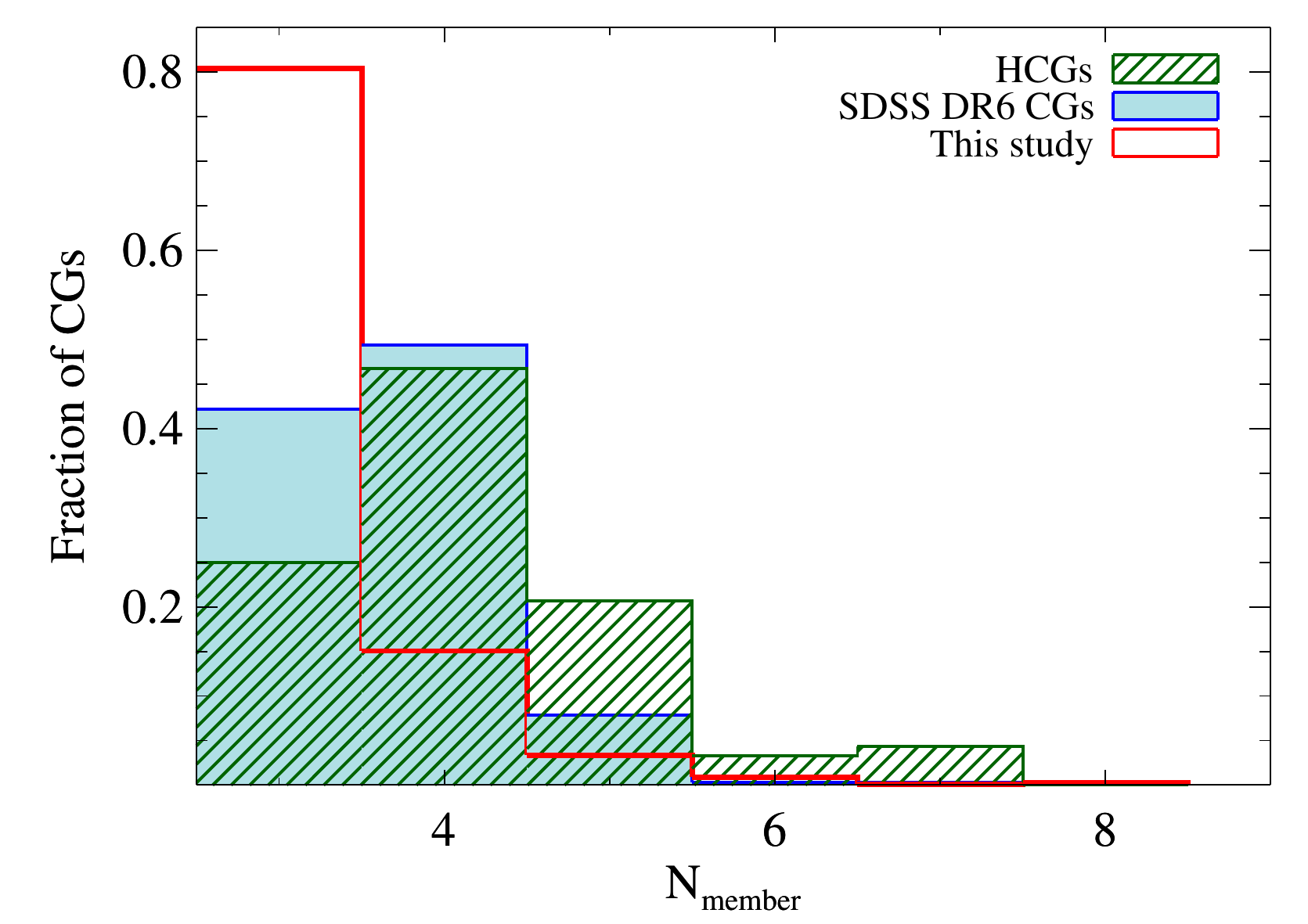}
\caption{
Population distribution of MLCG systems (open histogram) compared with 
 Hickson compact groups (hatched histogram) and
 SDSS DR6 compact groups (filled histogram).  } 
\label{num}
\end{figure}

The MLCG contains 1588 compact groups: 
 312 compact groups contain $N \geq 4$ members, 
 and 1276 group candidates contain $N=3$ members. 
There are more $N=3$ compact groups than $N\ge4$ compact groups
 at all redshifts.
Figure \ref{num} shows the distribution of the number of members in the MLCG systems.
Table \ref{grtab} lists the MLCG compact group candidates 
 including ID, R.A., Decl., the number of members, group redshift,
 group size, galaxy number density and rest frame line-of-sight velocity dispersion. 
We also note whether the group contains a tighter compact group
 satisfying the rest frame line-of-sight linking length $|\Delta V| \leq 500 \kms$, 
 V1CGs and V2CGs. 
Table \ref{galtab} lists the 5179 member galaxies contained in the compact groups of Table \ref{grtab}. 
We include
 the Group ID, Galaxy ID, R.A., Decl., morphology, $r-$band extinction- and $k-$corrected model magnitude, 
 $u-r$ color, stellar mass, redshift and its source. 

%%=================================
%%%%   Table 3
%%=================================
\begin{table*}
  \begin{center}
    \caption{Catalog of MLCG Members}
    \label{galtab}
    \begin{tabular}{cccccccccc}
    \hline
    Group ID & Galaxy ID\tablenote{SDSS DR12 object ID} & R.A. & Decl. & 
    Morph.\tablenote{Galaxy morphology. 1 indicate early types and 2 indicates late types.} &
    $r$\tablenote{The SDSS extinction- and $k-$corrected model magnitudes.} & 
    $u-r^{\rm c}$ & z & $\log (M_{stellar}/M_{\odot})$ &
    z source\tablenote{Source of the galaxy redshift.} \\
    \hline 
    MLCG0001 & 1237661387083284633 & 251.569153 &  31.726006 & 1 & 15.02 &  2.82 & $0.0541 \pm 0.00002$ & $10.761^{+0.116}_{-0.068}$ & SDSS \\
    MLCG0001 & 1237661387083284634 & 251.560486 &  31.725866 & 2 & 15.82 &  2.78 & $0.0522 \pm 0.00002$ & $10.403^{+0.239}_{-0.181}$ & SDSS \\
    MLCG0001 & 1237661387083284893 & 251.549835 &  31.714281 & 1 & 17.47 &  2.58 & $0.0538 \pm 0.00003$ & $ 9.568^{+0.249}_{-0.136}$ & SDSS \\
    MLCG0002 & 1237661383844036794 & 140.154343 &  33.706100 & 1 & 16.62 &  2.10 & $0.0224 \pm 0.00002$ & $ 9.145^{+0.152}_{-0.086}$ & SDSS \\
    MLCG0002 & 1237661383844102154 & 140.191086 &  33.704514 & 1 & 14.73 &  3.79 & $0.0231 \pm 0.00001$ & $10.233^{+0.129}_{-0.063}$ & SDSS \\
    MLCG0002 & 1237661383844102157 & 140.200195 &  33.679672 & 1 & 16.65 &  2.45 & $0.0224 \pm 0.00001$ & $ 9.016^{+0.256}_{-0.109}$ & SDSS \\
    MLCG0002 & 1237661383844037025 & 140.183090 &  33.654678 & 2 & 17.73 &  1.73 & $0.0236 \pm 0.00001$ & $ 8.416^{+0.236}_{-0.115}$ & SDSS	\\
    \hline
    \end{tabular}
  \end{center}
\tablecomments{
The full table is available in the online journal. A portion is shown here for guidance regarding its form and content.}
\end{table*}

%%%=================================
%%%%%   Table 4
%%%=================================

\begin{table*}
  \begin{center}
    \caption{Morphological Composition}
    \label{morph}
    \begin{tabular}{lcccc}
    \hline
    Catalog & CG type & $N_{galaxy}$ & Early types & Late types \\
    \hline
    \multirow{3}{*}{MLCGs} & Total     & 5179 & 3316 ($64.0 \pm 0.01\%$) & 1993 ($36.0 \pm 0.01\%$) \\
                           & $N \ge 4$ & 1351 &  943 ($69.8 \pm 0.01\%$) &  433 ($30.2 \pm 0.03\%$) \\
                           & $N = 3$   & 3828 & 2373 ($62.0 \pm 0.01\%$) & 1560 ($38.0 \pm 0.01\%$) \\
    \hline
	\multirow{3}{*}{V1CGs} & Total     & 2190 & 1503 ($68.6 \pm 0.01\%$) &  687 ($31.4 \pm 0.01\%$) \\
                       	   & $N \ge 4$ &  531 &  405 ($76.3 \pm 0.02\%$) &  126 ($23.7 \pm 0.02\%$) \\
                           & $N = 3$   & 1659 & 1098 ($66.2 \pm 0.01\%$) &  561 ($33.8 \pm 0.01\%$) \\
	\hline
	\multirow{3}{*}{V2CGs} & Total     &  930 &  704 ($75.7 \pm 0.01\%$) &  226 ($24.3 \pm 0.01\%$) \\
                           & $N \ge 4$ &  147 &  127 ($86.4 \pm 0.03\%$) &   20 ($13.6 \pm 0.03\%$) \\
                           & $N = 3$   &  783 &  577 ($73.7 \pm 0.02\%$) &  206 ($26.3 \pm 0.02\%$) \\
    \hline
    \end{tabular}
  \end{center}
\end{table*}

We also examine the morphological composition of
  the compact groups (Table \ref{morph}). 
The fraction of early-type galaxies in the MLCG 
  is $64.0 \pm 0.01\%$, exceeding the fraction in the Hickson compact groups 
 (i.e. $51 \pm 2 \%$, \citealp{Hic88}). 
The early-type fraction for $N\ge4$ compact groups ($69.8 \pm 0.01\%$) exceeds 
the fraction for $N=3$ compact groups ($62.0 \pm 0.01\%$). 
The error in the fraction of early-type galaxies is 
 the $1\sigma$ deviation from 1000 bootstrap resamplings.

\begin{figure}
\centering
\includegraphics[width=8.5cm,angle=0]{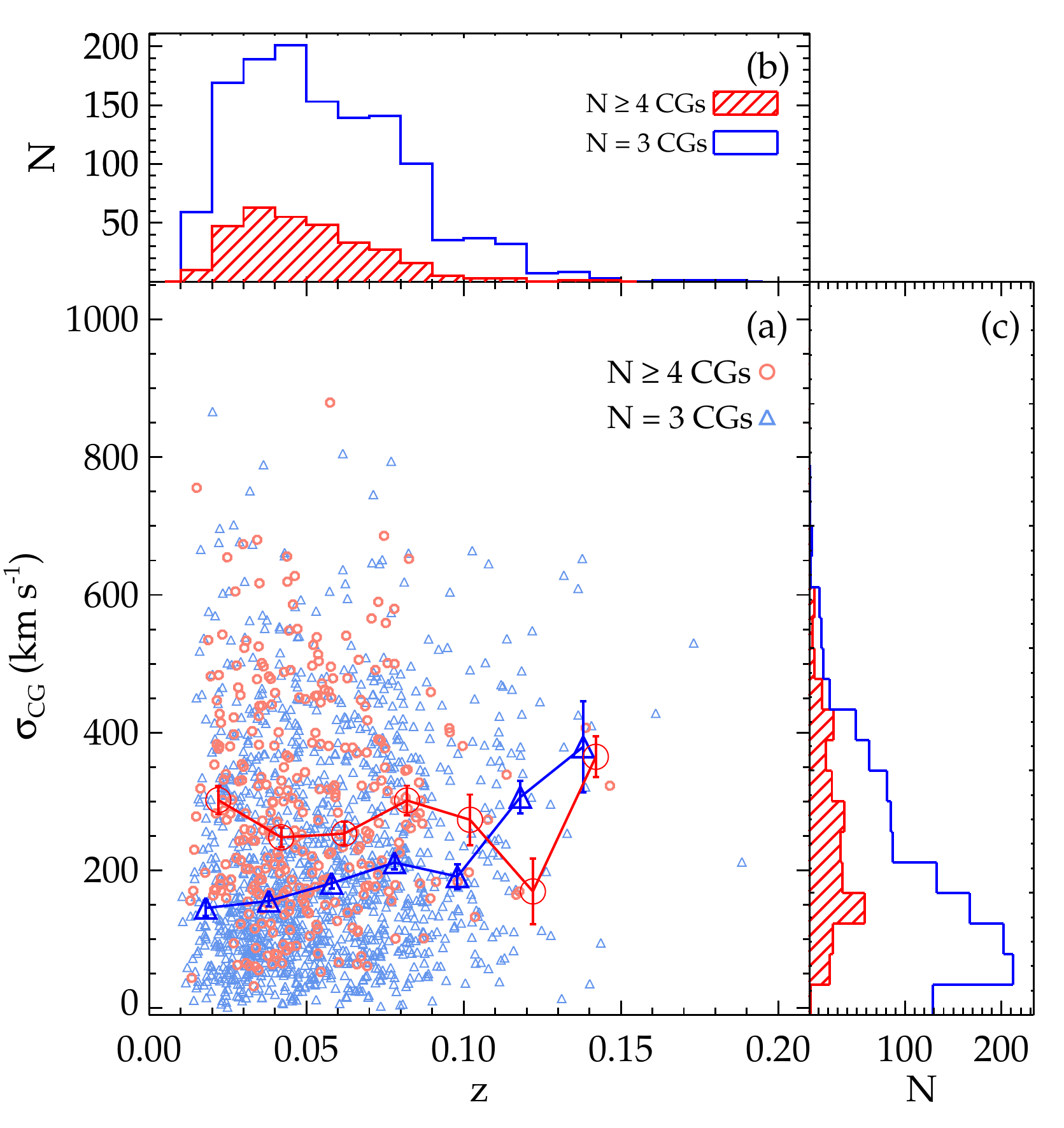}
\caption{
(a) Compact group velocity dispersion as a function of redshift. 
Open circles denote $N \ge 4$ compact groups; 
 triangles denote $N = 3$ compact groups. 
Larger symbols represent 
 the median velocity dispersion of MLCG systems in redshift bins $\Delta{z} =0.1$ .   
(b) Redshift distribution of compact groups
  for $N \ge 4$ (hatched) and $N = 3$ (open) compact groups.
(c) Histograms of the velocity dispersion of MLCG systems.
 The definitions of the histograms are the same as for panel (b) }
\label{ml_vol_grp}
\end{figure}

Figure \ref{ml_vol_grp} shows 
 the velocity dispersion of the MLCG compact groups 
 as a function of redshift in the range $0.01 < z < 0.19$. 
We calculate the rest frame line-of-sight velocity dispersion of each compact groups, 
 $\sigma_{CG}$, from equation (1) of \citet{Dan80}. 
As expected, the average velocity dispersion of $N\ge4$ compact group 
 is generally larger than for $N=3$ compact groups. 

\begin{figure}
\centering
\includegraphics[width=8.5cm,angle=0]{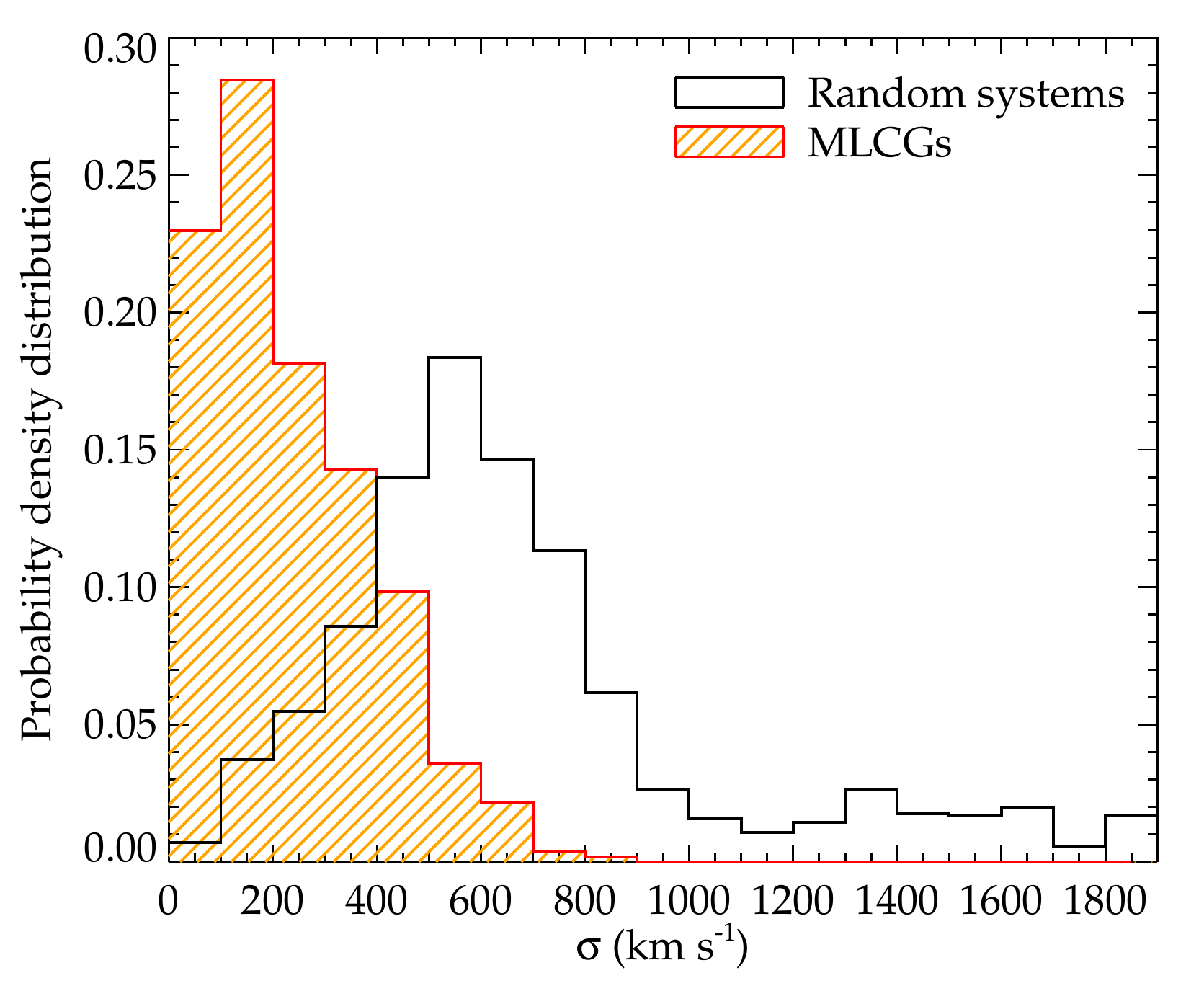}
\caption{
Probability distribution function for 
 $\sigma_{CG}$ (hatched histograms) and 
 $\sigma_{los}$ (open histograms). }
\label{prob}
\end{figure}

There are 256 compact groups
 with very large velocity dispersion $\sigma_{CG} \geq 400 \kms$,
 overlapping the distribution for galaxy clusters ($ \sim 400 - 1300 \kms$, \citealp{Rin13}).
Compact group candidates with large $\sigma_{CG}$ also appear in other catalogs 
 (e.g. \citealp{Bar96, Pom12, Soh15}). 
To examine the possibility that 
 these higher velocity dispersion groups are merely superpositions along the line-of-sight,
 we compare the velocity dispersion distribution of the MLCG systems with the distribution for 
 a set of $\sim3000$ randomly selected triplets and quadruplets 
 in the same redshift and apparent magnitude range.
To construct these random superpositions, 
 we first pick a galaxy at a randomly chosen MLCG redshift and then 
 randomly select two or three other galaxies 
 within $|\Delta V| \leq 1000 \kms$ without attention to the projected spatial separation. 
The velocity dispersion distribution of these random samples is 
 broad with $40 \kms < \sigma_{los} < 2300 \kms$
 with a median value of $\sigma_{los} \sim 604 \kms$, 
 a factor of three larger than for the MLCG objects. 
Figure \ref{prob} shows the probability distribution of the MLCG velocity dispersions (red) 
 along with the distributions for the fictitious groups 
 randomly selected from the survey redshift distribution. 
The distribution for the fictitious distribution (black) is 
 appropriately weighted for the $N=3$ and $N\geq4$ samples. 
Both the Kolmogorov-Smirnov and the Anderson-Darling k-sample tests reject the hypothesis that 
 these distributions are derived from the same parent distribution ($p = 0$).
The overlap of the two distributions suggest that 
 MLCG systems with $\sigma_{CG} \geq 500~\kms$ may often contain superpositions. 
Furthermore, we can estimate an upper limit 
 on the fraction of MLCG systems with $\sigma_{CG} \lesssim 400 \kms$
 that may be contaminated by interlopers by computing the 
 integral of the product of the distributions in the region of overlap. 
The limit is about 18\%. 
Obviously the probability that a candidate compact group includes 
 a superposition increases as the velocity dispersion increases.

\begin{figure}
\centering
\includegraphics[width=8.5cm,angle=0]{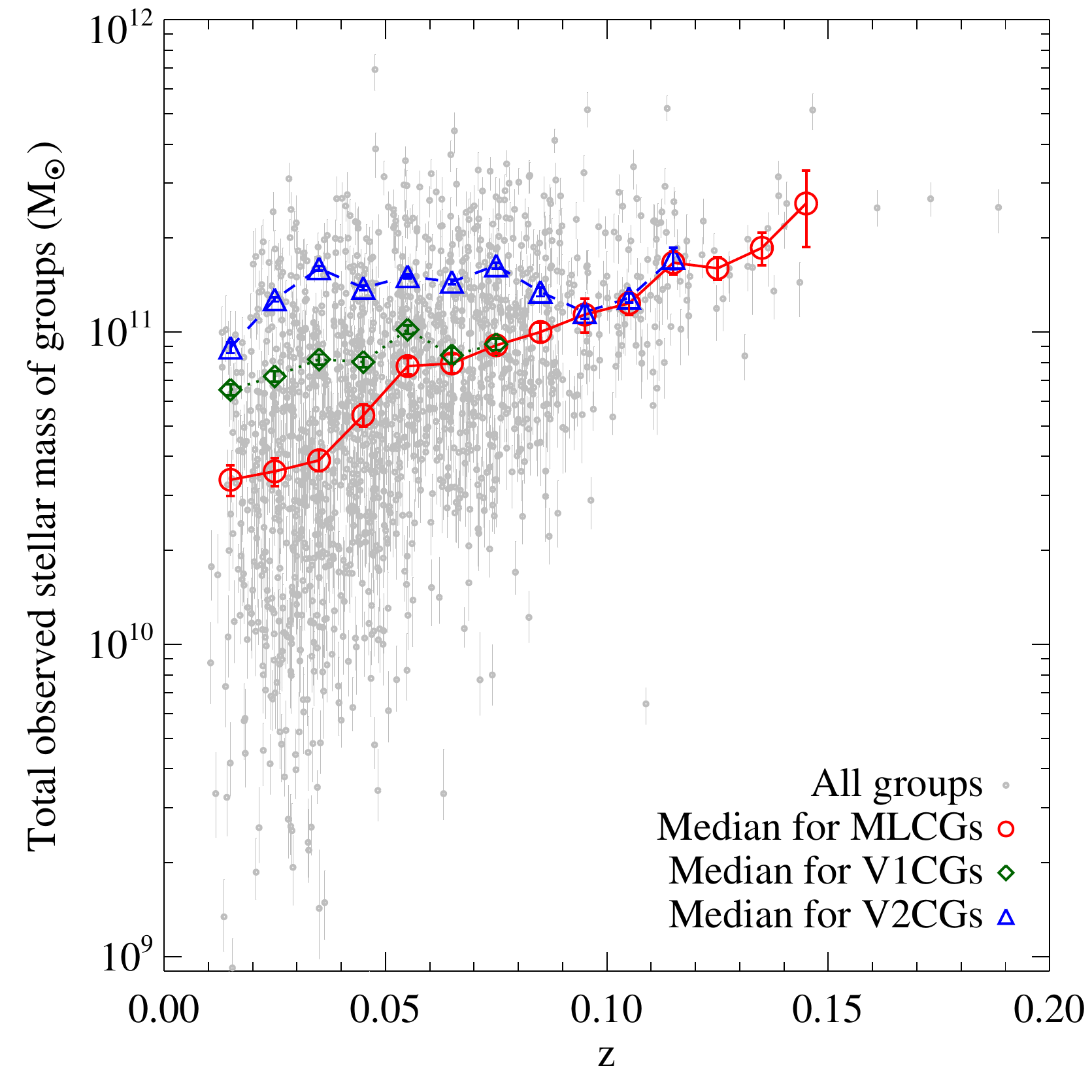}
\caption{
Observed stellar mass of MLCG systems as a function of redshift (small points) 
 and their median in redshift bins.
The median of observed group stellar mass among the
 V1CGs (diamonds) and the V2CGs (triangles) can be compared directly 
 at different redshifts because they require no relative correction 
 for the unobserved end of the stellar mass function. }
\label{mlzmass}
\end{figure}

Figure \ref{mlzmass} shows the 
 total observed stellar masses of the compact groups in the MLCG
 as a function of redshift. 
The stellar masses range from 
 $9.2 \times 10^{8}~M_{\odot} < M_{stellar} < 6.9 \times 10^{11}~M_{\odot}$,
 similar to compact groups in the SDSS DR6 \citep{Coe15}. 
The median observed total stellar mass of the MLCG systems 
 increases as the group redshift increases. 
This dependence results from the nature of the parent magnitude-limited sample. 
 
\subsubsection{\label{vol}Volume-Limited Subsamples of the MLCG}

Volume-limited subsamples have the advantage that 
 the physical properties of the candidate systems should be the same throughout the volume. 
Each volume-limited sample also contains systems 
 with a narrow range of total stellar mass. 
The two volume-limited samples we select highlight compact systems 
 with stellar masses in the ranges:
 $6.1 \times 10^{9}~M_{\odot} < M_{MLCG} < 6.9 \times 10^{11}~M_{\odot}$ (V1CG) and 
 $6.5 \times 10^{9}~M_{\odot} < M_{MLCG} < 5.2 \times 10^{11}~M_{\odot}$ (V2CG). 

The sample V1CG contains 675 groups in the redshift range $0.01 < z < 0.076$ (Table 5).%\ref{v1gr}).
These groups are similar to the median groups in the MLCG for the redshift range $0.06 < z < 0.08$.
In contrast with the MLCG systems (red points in Figure \ref{mlzmass}), 
 the median stellar mass for the V1CG groups is essentially constant throughout the sample redshift range. 
In Table \ref{grtab} we indicate the MLCG groups that are also contained in the V1CG. 

V2CG contains 297 groups in the redshift range $0.01 < z < 0.116$ 
 with larger total stellar masses (Table 6). % \ref{v2gr}).  
For $z \lesssim 0.02$ the median total stellar mass drops 
 because the survey volume is too small to contain many of the most massive objects.
For $z \geq 0.02$, 
 the median total stellar mass of the systems is approximately constant 
 throughout the redshift range. 
The median matches the median for the MLCG for redshifts greater than $z > 0.09$. 
The V2CG systems are a subset of both the V1CG systems and the MLCG systems. 
Table \ref{grtab} indicates group membership in these subsamples.

In Section \ref{vlsam} we use the V1CG and V2CG subsets of the MLCG
 to examine the properties of compact group candidates 
 of the same stellar mass as a function of redshift. 
We also discuss use of these samples 
 as a basis for discussing the impact of the SDSS fiber-positioning constraints on the 
 identification of compact group candidates.
 
%%=================================
%%%%   Table 5
%%=================================
\begin{table*}
 \begin{center}
  \caption{Catalog of V1CGs}
  \begin{tabular}{ccccccccc}
  \hline
  \multirow{2}{*}{ID\tablenote{Member galaxies are contained in the MLCG galaxy catalog (Table \ref{galtab}).}} & 
  R.A. & Decl. & \multirow{2}{*}{$N_{mem}$} & 
  \multirow{2}{*}{z\tablenote{The error is the $1\sigma$ deviation derived from 1000-times bootstrap resamplings.}} & 
  $R_{gr}^{\rm b}$  & $\log \rho^{\rm b}$ & $\sigma_{CG}^{\rm b}$ & 
  \multirow{2}{*}{$N_{C}$\tablenote{The number of neighboring galaxies in the comoving cylinder.}} \\  
     & (J2000) & (J2000) &           &   & ($h^{-1}$ kpc) & ($h^{3}$ Mpc$^{-3}$) & ($\kms$)      &         \\ 
  \hline
  V1CG001 & 198.227173 &   1.012775 &  3 & $0.0723 \pm 0.0008$ & $ 37.8 \pm  5.4$ & $4.12 \pm 0.32$ & $ 482 \pm 116$ &   8 \\
  V1CG002 & 139.935760 &  33.744308 &  3 & $0.0202 \pm 0.0012$ & $ 13.0 \pm  1.5$ & $5.52 \pm 0.19$ & $ 866 \pm 207$ &   2 \\
  V1CG003 & 154.741531 &  37.298065 &  3 & $0.0480 \pm 0.0003$ & $ 32.9 \pm  5.5$ & $4.30 \pm 0.41$ & $ 210 \pm  37$ &   1 \\
  V1CG004 & 158.222275 &  12.086633 &  3 & $0.0330 \pm 0.0004$ & $ 12.3 \pm  3.0$ & $5.58 \pm 0.68$ & $ 242 \pm  68$ &   5 \\
  V1CG005 & 127.709404 &  28.573534 &  3 & $0.0657 \pm 0.0000$ & $ 31.2 \pm  3.7$ & $4.37 \pm 0.25$ & $  26 \pm   4$ &   1 \\
  \hline
  \end{tabular}
 \end{center}
 \label{v1gr}
 \tablecomments{
  The full table is available in the online journal. A portion is shown here for guidance regarding its form and content.}
\end{table*}

%%=================================
%%%%   Table 6
%%=================================
\begin{table*}
 \begin{center}
  \label{v2gr}
  \caption{Catalog of V2CGs}
  \begin{tabular}{ccccccccc}
  \hline
  \multirow{2}{*}{ID\tablenote{Member galaxies are contained in the MLCG galaxy catalog (Table \ref{galtab}).}} & 
  R.A. & Decl. & \multirow{2}{*}{$N_{mem}$} & 
  \multirow{2}{*}{z\tablenote{The error is the $1\sigma$ deviation derived from 1000-times bootstrap resamplings.}} & 
  $R_{gr}^{\rm b}$ & $\log \rho^{\rm b}$ & $\sigma_{CG}^{\rm b}$ & 
  \multirow{2}{*}{$N_{C}$\tablenote{The number of neighboring galaxies in the comoving cylinder.}} \\
     & (J2000) & (J2000) &           &   & ($h^{-1}$ kpc) & ($h^{3}$ Mpc$^{-3}$) & ($\kms$)      &         \\ 
  \hline
   V2CG001 & 154.930054 &  37.472149 &  3 & $0.0933 \pm 0.0002$ & $ 35.7 \pm  1.8$ & $4.20 \pm 0.09$ & $ 144 \pm  35$ &   1 \\
   V2CG002 & 142.154663 &  36.477848 &  3 & $0.0862 \pm 0.0003$ & $ 35.8 \pm  4.9$ & $4.19 \pm 0.36$ & $ 183 \pm  42$ &   0 \\
   V2CG003 & 206.682205 &  45.697647 &  4 & $0.0648 \pm 0.0002$ & $ 34.6 \pm  2.9$ & $4.36 \pm 0.13$ & $ 128 \pm  30$ &   8 \\
   V2CG004 & 210.859863 &  41.869808 &  3 & $0.1130 \pm 0.0001$ & $ 40.8 \pm  4.5$ & $4.02 \pm 0.22$ & $  73 \pm  17$ &   0 \\
   V2CG005 & 179.164658 &  11.389394 &  3 & $0.0682 \pm 0.0005$ & $ 33.2 \pm  5.2$ & $4.29 \pm 0.33$ & $ 357 \pm 113$ &   5 \\
   \hline
  \end{tabular}
 \end{center}
 \tablecomments{
 The full table is available in the online journal. A portion is shown here for guidance regarding its form and content.}
\end{table*}

\subsection{Comparison of the MLCG with Photometrically Selected Samples}\label{m09}

The SDSS DR6 compact group candidate sample of \citet{McC09} 
 is the largest catalog previously available; 
 it includes
   2297 compact group candidates drawn from a magnitude-limited sample with 
 $14.5 \leq r \leq 18.0$ (catalog A) and 
 74,791 compact group candidates from a magnitude-limited sample with 
 $14.5 \leq r \leq 21.0$ (catalog B). 
\citet{McC09} identified these compact group candidates 
 by applying all of Hickson's criteria to the SDSS DR6 photometric galaxy sample. 

Because the primary identification of the \citet{McC09} groups is photometric, 
 the interloper fraction is substantial. 
By measuring additional redshifts, 
 \citet{Soh15} estimate that the fraction is greater than 40\%. 
\citet{Men11} pruned the \citet{McC09} catalog by 
 using photometric redshifts. 
However, the uncertainty in photometric redshifts (median $\sim2800\kms$) is large 
 compared with the typical velocity separation among candidate group member galaxies. 
In spite of these limitations, the \citet{McC09}  
 SDSS DR6 compact group candidate sample provides 
 a basis for comparison with the MLCG. 
The comparison tests
 the impact of different group selection methods.  
 
We match the MLCG compact groups 
 with group candidates in catalogs A and B of \citet{McC09}
 based on angular separation. 
We count the number of MLCG systems
 matched ($D_{sep} < R_{gr}$) with group candidates in either catalog,
 where $D_{sep}$ is the angular separation between the MLCG system and a \citet{McC09} group candidate,
 and $R_{gr}$ is the angular radius of the MLCG. % sample compact groups. 
Only 242 (15\%) of the MLCG systems overlap with
 compact group candidates identified by \citet{McC09}. 
This low matching rate results primarily 
 from (1) the MLCG inclusion of galaxies with $r<14.5$ and 
 from (2) differences in the group identification algorithm.
 
We next examine the reasons that individual MLCG systems
 are missing from the \citet{McC09} catalog in more detail.   
First, 
 there are 384 (24\%) groups in the MLCG that
 contain bright ($r < 14.5$) member galaxies 
 excluded from the input galaxy catalog. 
Most of these groups are located at $z < 0.05$
 where \citet{McC09} identified only a few compact group candidates \citep{Soh15}.
Second,
 1062 (67\%) MLCG systems
 do not satisfy the isolation criterion originally applied by \citet{Hic82} 
 and followed by \citet{McC09}. 
These groups have one or more non-member galaxies ($\Delta r < 3$ mag) 
 within the isolation annulus ($R_{gr} < R_{GCD} < 3 R_{gr}$),
 where $R_{GCD}$ is the groupcentric distance and $R_{gr}$ is the projected group radius
 (see Section \ref{iso} for further discussion). 
Third, 472 (30\%) triplets in the MLCG 
 satisfy all of the \citet{McC09} group candidate selection criteria except that 
 there are only three members with accordant redshifts.   
The total number of compact groups 
 that violate each of the selection criteria of \citet{McC09}
 exceeds the total number of MLCG systems
 because some MLCG groups violate more than one criterion.

\begin{figure}
\centering
\includegraphics[width=8.5cm, angle=0]{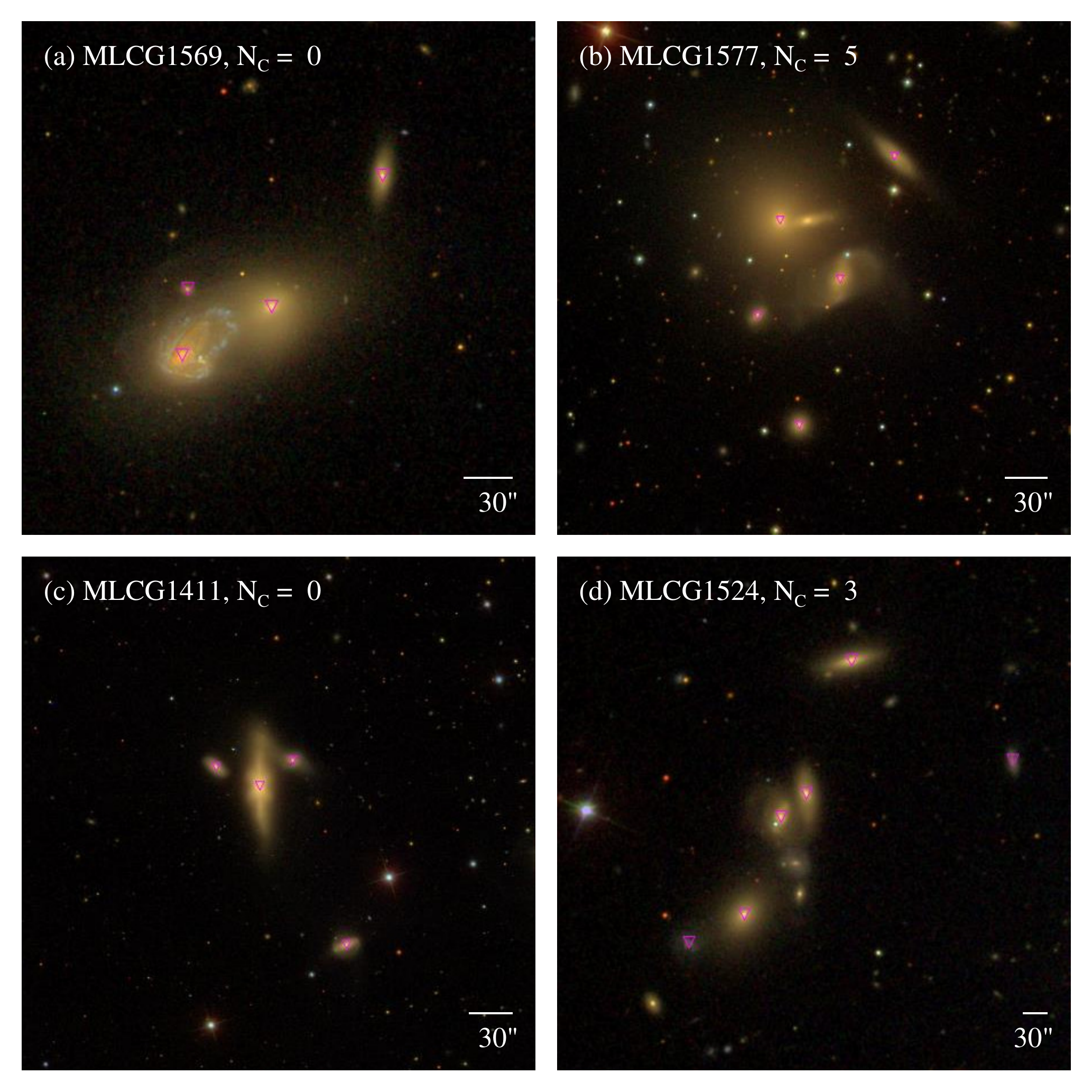}
\caption{
Sample images of MLCG systems absent
 from previous compact group catalogs. 
All these candidates violate the isolation criterion. 
Furthermore, 
 (a) MLCG1569, (b) MLCG1577, and (c) MLCG1411 are missing 
 because they contain bright galaxy ($r<14.5$).
Note the striking evidence for tidal interactions in all four systems. }
\label{ex}
\end{figure}

Figure \ref{ex} shows examples of MLCG systems
 absent from previous catalogs. 
These groups either contain bright members with $r < 14.5$ or they
 violate the isolation criterion. 
The member galaxies show active interacting features indicating that 
 they are physically bound systems
 rather than superpositions of galaxies along the line-of-sight.  
These examples underscore the impact of deriving a compact group catalog 
 from a complete redshift survey. 
%Hickson applied his isolation criterion because redshifts were scarce. 
%With a large survey like SDSS, this approach is no longer necessary. 

%=============================================================
\section{COMPACT GROUP PROPERTIES}

\subsection{\label{prop}Comparison with Other Compact Groups}

\begin{figure}
\centering
\includegraphics[width=8.5cm,angle=0]{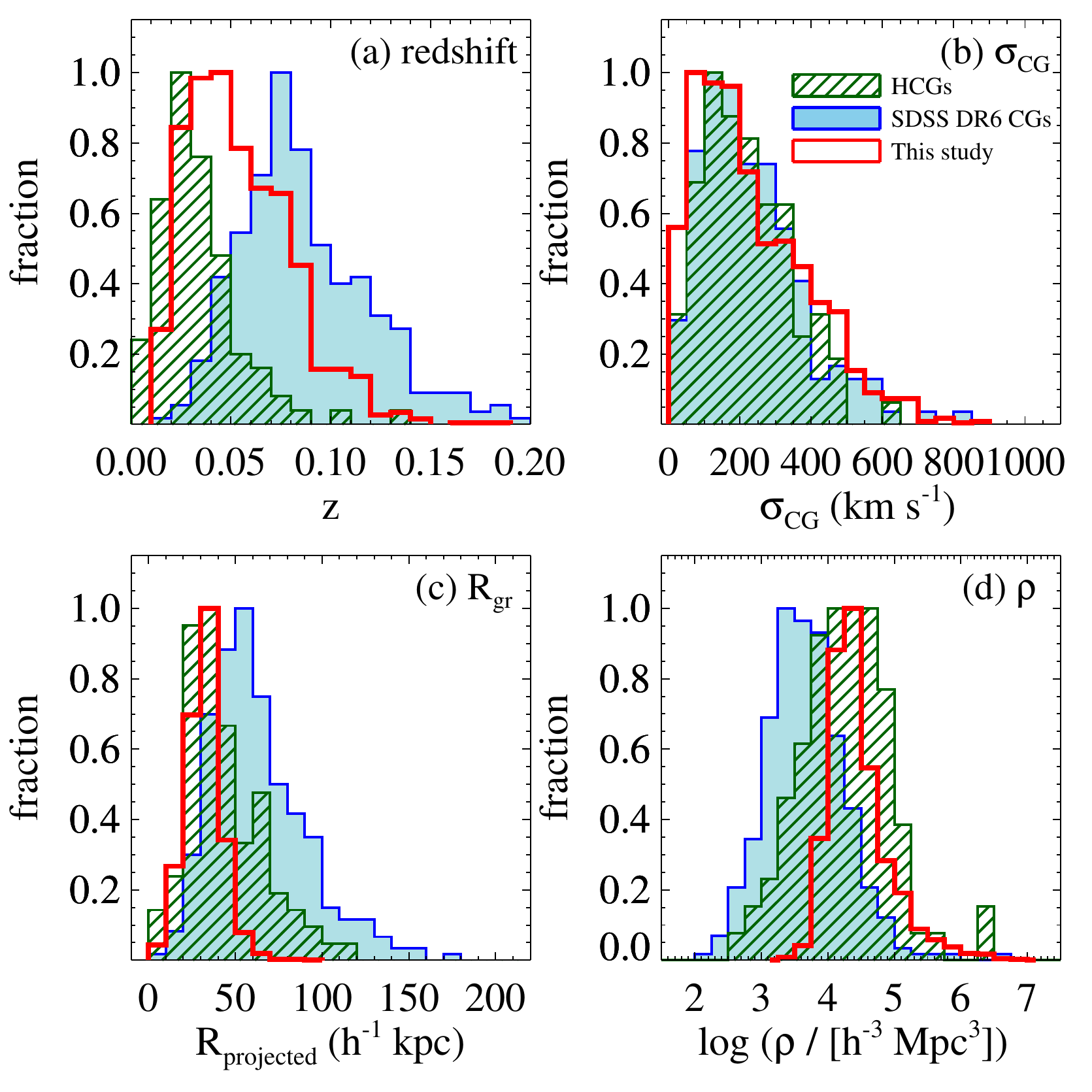}
\caption{
Properties of MLCG systems
 including (a) redshift, (b) velocity dispersion, 
 (c) projected group radius and (d) the galaxy number density
 compared with other compact group catalogs. 
The color and fill of the histograms are 
 the same as in Figure \ref{num}. }
\label{comp}
\end{figure}

We compare the physical properties of the MLCG systems
 with the Hickson compact groups \citep{Hic92} and 
 with the SDSS DR6 compact groups \citep{McC09, Soh15}. 
The SDSS DR6 compact groups we consider (SDSS DR6 compact groups hereafter)
 have spectroscopic redshifts
 from our FLWO/FAST observations and the literature, 
 including the SDSS DR12 \citep{Soh15}.
The identification of these systems was adapted from Hickson's criteria 
 (catalog A of \citealp{McC09}).
Thus, this comparison provides a measure of the way
 apparent compact group properties might depend on the selection method. 

Figure \ref{comp} compares
 the distributions of physical properties of the compact groups
 including redshift, velocity dispersion, projected group radius and number density.  
The redshift range of the MLCG systems, $0.01 < z < 0.19$, covers 
 the range for the Hickson compact groups,
 but differs from the range for the SDSS DR6 compact groups
 ($0.03 < z < 0.20$ with a median $z \sim 0.08$). 
Because we include bright galaxies and because we do not apply an isolation criterion,  
 we find more compact groups at $z < 0.03$ (See Section \ref{iso}). 
We miss compact group candidates at $z > 0.09$ 
 because of the SDSS incompleteness (see Section \ref{cgsam}) and magnitude limit. 
 
The MLCG systems have rest frame velocity dispersions $\sigma_{CG}$
  $\leq 880~\kms$ with a median of $194 \pm 4~\kms$,
 similar to that for other samples. 
For example, the median velocity dispersions are 
 $204 \pm 13~\kms$ for the Hickson compact groups, and 
 $207 \pm 31~\kms$ for the SDSS DR6 compact groups. 
The similar selection limit for 
 the radial separation between member galaxies 
 (i.e., $|\Delta V| < 1000 \kms$),
 essentially dictates that the velocity dispersion of the samples be similar.

The projected sizes of the MLCG systems
 differ from those of groups in other catalogs. 
The MLCG systems have $R_{\rm gr}$
 ranging from 4 to $94~h^{-1}$ kpc. 
The median $R_{\rm gr}$ = $32.0 \pm 0.3~h^{-1}$ kpc 
 is similar to that for the Hickson compact groups, 
 $38.6 \pm 7.1~h^{-1}$ kpc.
However, the median size of the SDSS DR6 compact groups, 
 $57.8 \pm 1.5~h^{-1}$ kpc,
 is much larger than for the MLCG systems or for the Hickson compact groups 
 \citep{McC09, Soh15}. 

Because the SDSS DR6 compact groups are 
 apparently larger than the compact groups identified in other catalogs, 
 the resulting galaxy number density appears to be smaller. 
The galaxy number density is  
\begin{equation}
\rho = \frac{3N}{4\pi R_{gr}^{3}},
\end{equation}
 where $N$ is the number of members and 
 $R_{gr}$ is the projected group radius in $h^{-1}$ Mpc.
The median galaxy number density of the MLCG systems is 
 $\log (\rho / [h^{-3}~{\rm Mpc}^3]) = 4.36 \pm 0.01$. 
The median is
 $\log (\rho / [h^{-3}~{\rm Mpc}^3]) = 4.27 \pm 0.09$ for the Hickson compact groups and
 $\log (\rho / [h^{-3}~{\rm Mpc}^3]) = 3.65 \pm 0.04$ for the SDSS DR6 compact groups. 
In other words,
 the number density of the MLCG systems is similar to that of the Hickson compact groups,
 but higher than that of the SDSS DR6 compact groups
  (bottom right panel of Figure \ref{comp}).

\begin{figure}
\centering
\includegraphics[width=8.5cm,angle=0]{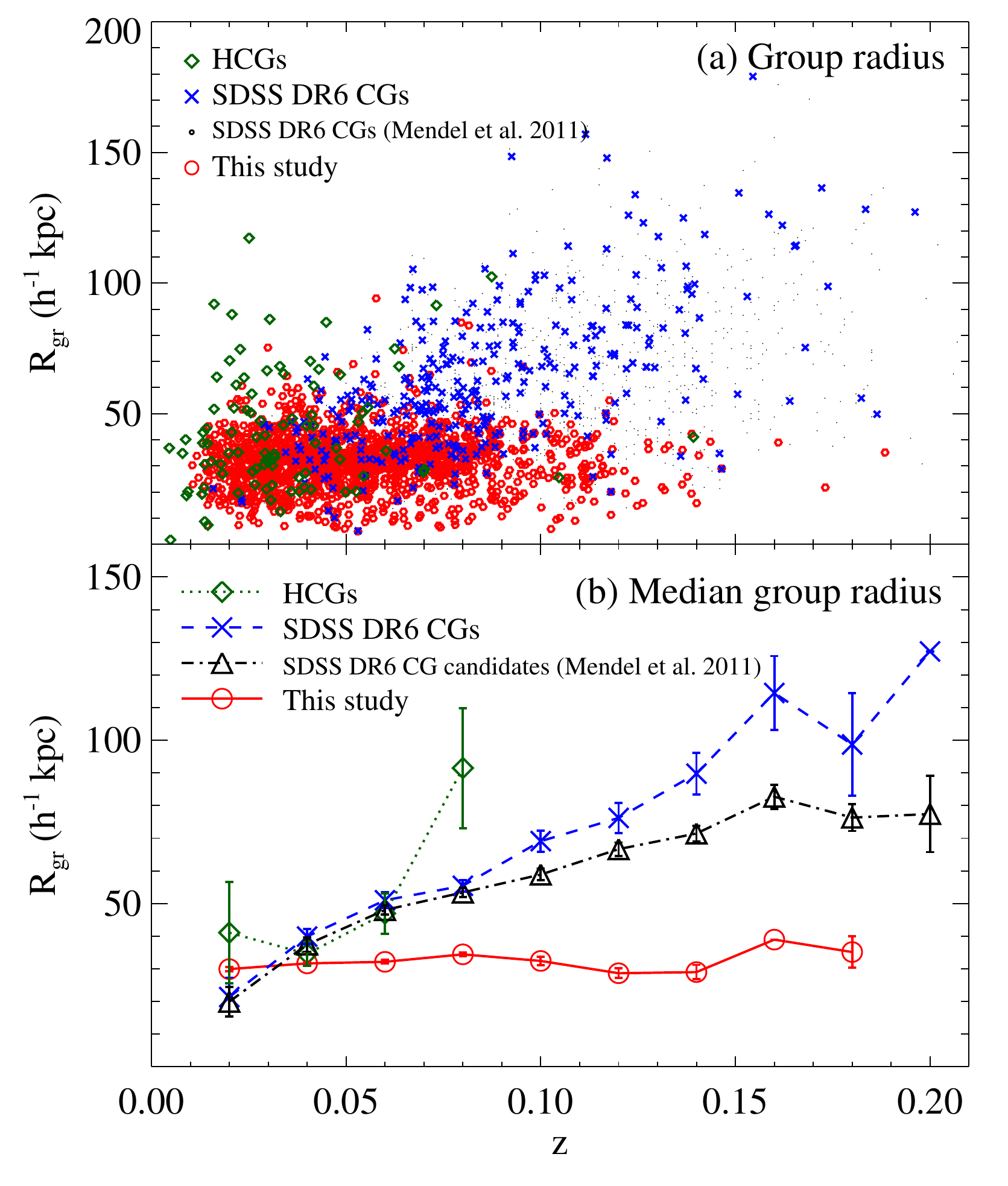}
\caption{
Projected sizes of the MLCG systems as a function of redshift.
Open circles, diamonds, triangles and crosses are 
  the MLCG systems, the Hickson compact groups, 
  the SDSS DR6 compact group candidates identified with photometric redshifts \citep{Men11} and
  the SDSS DR6 compact groups, respectively. Note the near constancy of the MLCG sizes.}
\label{mlzphysiz}
\end{figure}

In principle, 
 over the redshift range we explore here, 
 the physical properties of the compact group candidates 
 identified by an algorithm should not depend strongly on redshift.
In Figure \ref{mlzphysiz}, we compare properties of the catalogs 
 as a function of redshift beginning with the $R_{\rm gr}$. 
The median $R_{\rm gr}$ of the MLCG systems
 varies little with redshift;
 in contrast the sizes of compact groups in other samples  
 increase significantly with redshift.    
One possibility is that 
 \citet{Soh15} selected larger groups for their follow-up redshift measurements.  
However, 
 the median $R_{\rm gr}$ of compact group candidates
 based on photometric redshifts \citep{Men11} shows a similar trend. 
Thus we suspect that
 the trend in group size is related to some aspect of Hickson's criteria. 
For example a relatively nearby group with a large physical size 
 has a correspondingly large isolation region 
 thus increasing the probability that 
 the group candidate will be removed from the sample, 
 because an interloper appears in the isolation region. 

Because the median group size in previous samples varies with redshift, 
 the galaxy number density obviously also varies;
 the galaxy number density tends to be lower at higher redshift
 compared to the MLCG systems. 
The lower density at greater redshift 
 increases the probability that the group candidate contains interlopers and 
 decreases the probability of selecting a very dense system 
 where galaxy-galaxy interactions are likely. 
In other words, {\it intrinsic systematics in the selection 
 potentially may lead to artificially biased physical conclusions
 about the properties of the candidate systems as a function of redshift.}

\subsection{The Environment of Compact Groups}

\begin{figure}
\centering
\includegraphics[width=8.5cm, angle=0]{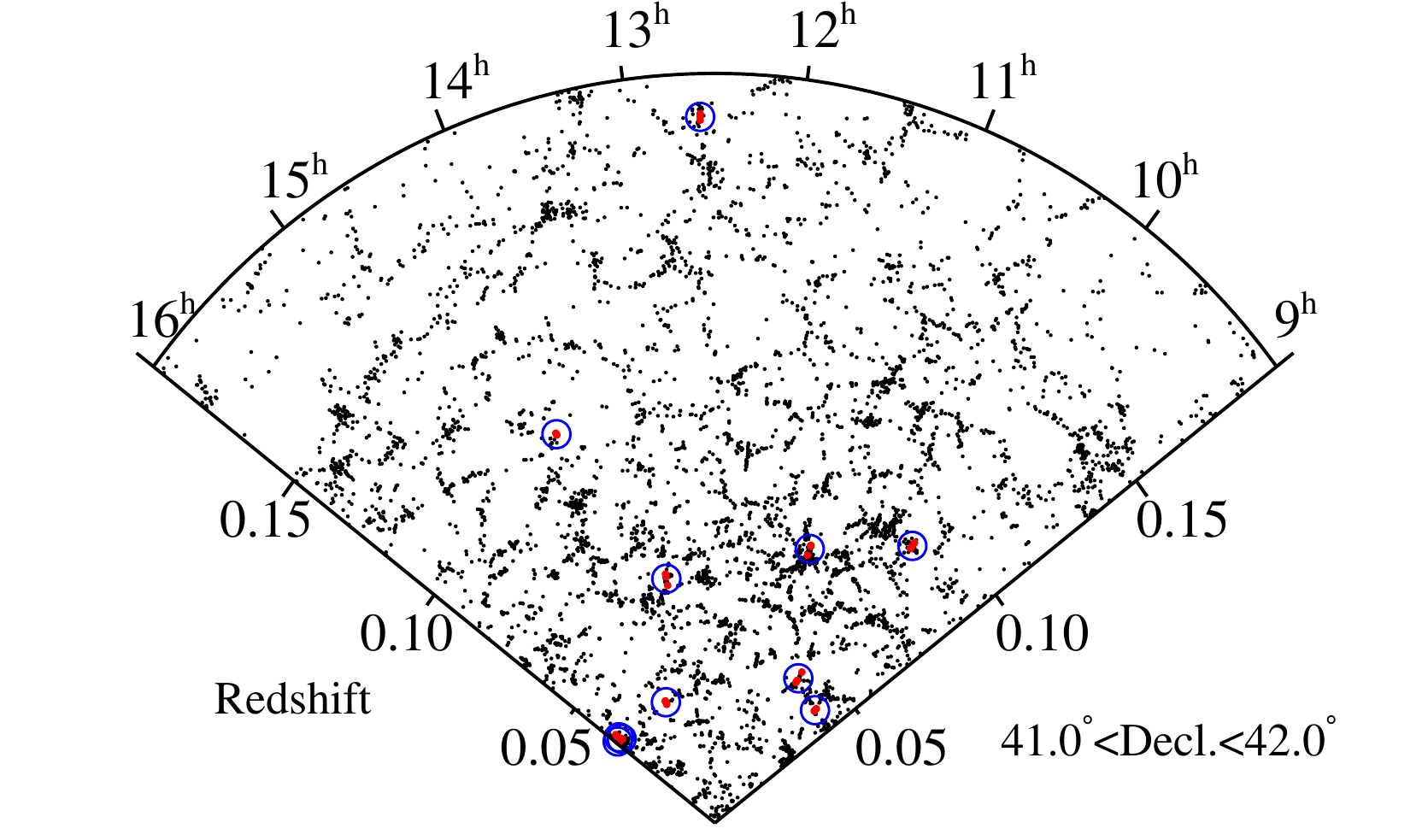}
\caption{
Example cone diagram for a slice with
 $9h < R.A. < 16h, ~41.0 ^{\circ}  < Decl. < 42.0 ^{\circ}$,
 and $0.0 < z < 0.2$.
Blue large and red small circles indicate
 compact groups and their member galaxies, respectively. 
Black dots denote SDSS galaxies with $M_{r} < -20.5$. }
\label{mlcone}
\end{figure}

Figure \ref{mlcone} shows an example cone diagram
 indicating the locations of some of the MLCG systems.
The points are the SDSS galaxies in the magnitude-limited sample. 
The MLCG systems reside in diverse environments,
 consistent with results from previous studies
 \citep{Ram94, Bar96, Rib98, Men11, Pom12, Dia15}. 

To study the local environments of compact groups quantitatively, 
 we define the number of neighboring galaxies ($N_{C}$) around each compact group
 within a comoving cylinder of $\Delta R < 700~h^{-1}$ kpc and rest frame $|\Delta V| < 1000 \kms$
 \citep{Bar07, Woo10} centered on the group mean position and redshift. 
We count $N_{C}$ for 1534 compact groups within a volume-limited sample 
 with $M_{r} < -20.0$ and $0.01 < z < 0.115$
 extracted from the SDSS DR12 spectroscopic sample 
 (blue box in Figure \ref{volume}). 
We exclude compact group member galaxies from the $N_{C}$ count.  
We also estimate $N_{C}$ for 
 254 SDSS DR6 compact groups (out of 332 groups) for comparison. 

\begin{figure}
\centering
\includegraphics[width=8.5cm]{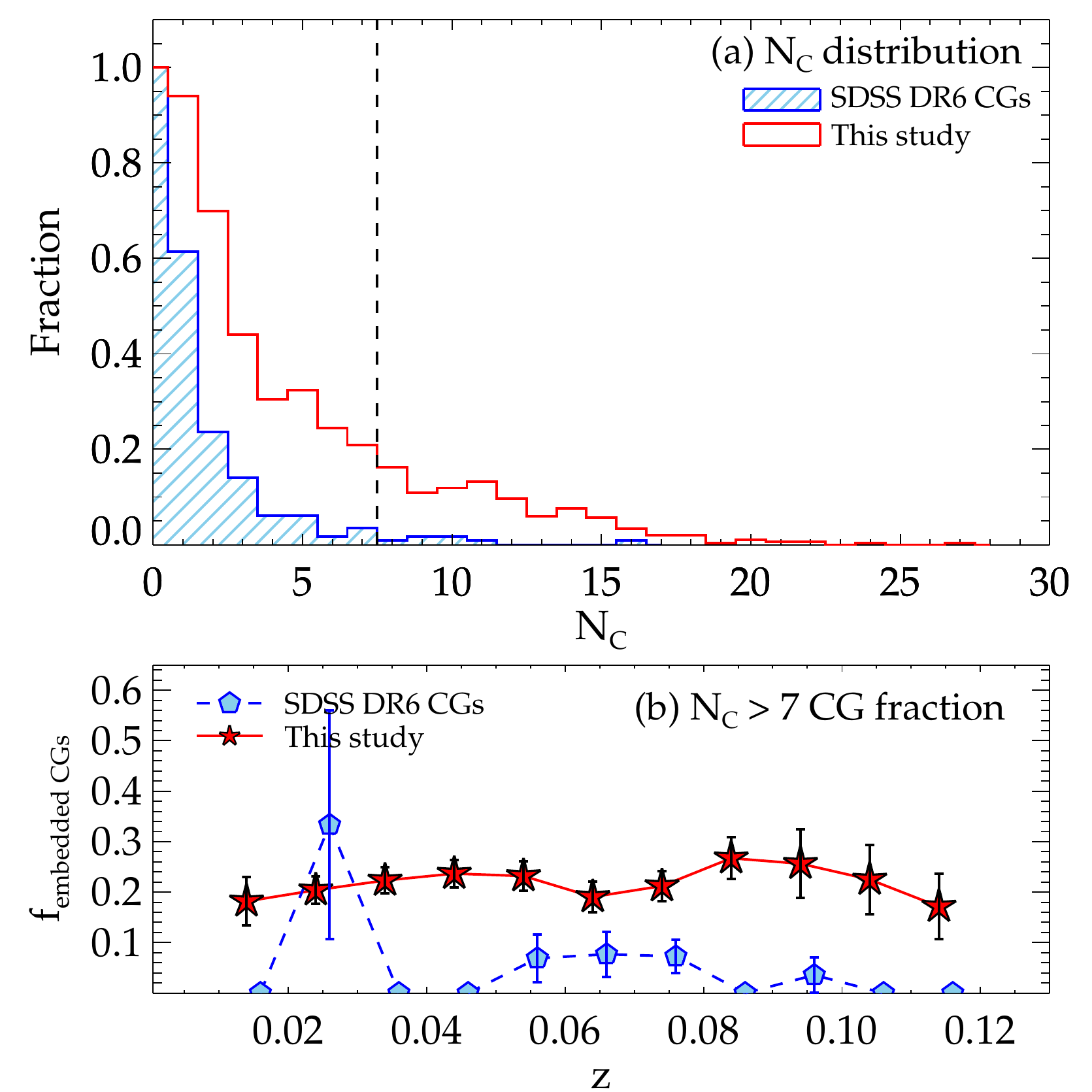}
\caption{
(a) Distribution of the number of neighboring galaxies ($N_{C}$)
 in a comoving cylindrical volume with $\Delta R < 700~h^{-1}$ kpc and $|\Delta V| < 1000 \kms$
 for the MLCG systems (open histogram) and 
 for the SDSS DR6 compact groups (hatched histogram). 
(b) The fraction of compact groups in dense environments ($N_{C} > 7$)
 as a function of redshift:
 stars indicate MLCG systems and 
 pentagons indicate SDSS DR6 compact groups. }
\label{nc}
\end{figure}

Figure \ref{nc} displays the $N_{C}$ distribution 
 for the MLCG systems and for the SDSS DR6 compact groups. 
Both group samples show a similar range of $N_{C}$, 
 but the distributions differ. 
There are more MLCG systems with larger numbers of neighbors. 
This difference results from
 the absence of an isolation criterion in the MLCG selection algorithm.

$N_{C}$ has been used as an environment measure 
 in both theoretical and observational studies of
 tight galaxy pairs.
For example, in a theoretical investigation,
 \citet{Bar07} segregated local environments of galaxy pairs at $N_{C} = 8$ and 
 \citet{Woo10} followed the procedure in the interpretation of observations.
\citet{Woo10} estimated that 32\% of galaxy pairs 
 are located in dense environments with $N_{C} > 8$. 
When they computed $N_{C}$, 
 they included a pair member galaxy for counting $N_{C}$; 
 in contrast we exclude the compact group member galaxies. 
Thus, $N_{C} =8$ in their studies corresponds to $N_{C}=7$.   
Among MLCG systems, 18\% are in denser regions ($N_{C} > 7$), 
 apparently somewhat lower than the fraction for galaxy pairs. 
However our lower number may result from the SDSS undersampling of dense regions; 
 this issue is much less important for pair samples from \citet{Woo10}.  
%\citet{Bar07} based their catalog on the 2dF galaxy redshift survey \citep{Col01} 
% which is unaffected by any significant undersampling of dense regions. 
\citet{Woo10} constructed pair catalogs based on the SHELS survey \citep{Gel05, Gel14}. 
The catalog for this survey is 97\% complete to the survey limit and 
 thus the biases are negligible.
    
The bottom panel of Figure \ref{nc} shows the fraction of compact groups in the denser environments
 as a function of redshift.
The fraction changes little in the redshift range $0.02 \leq z \leq 0.12$. 
The fraction of MLCG systems in denser environments 
 always exceeds that for the SDSS DR6 groups 
 as a result of the differences in the group identification algorithm. 

%=============================================================
\section{SELECTION ISSUES}

\subsection{\label{iso} Isolation Criteria}

In his identification of compact groups, 
 Hickson applied an isolation criterion 
 to compensate for observational limitations and 
 to avoid systems embedded in massive clusters \citep{Hic82}. 
\citet{Bar96} pointed out that 
 the availability of large complete redshifts surveys obviates 
 the need for applying an isolation criterion 
 in the initial selection of compact group candidates. 
They emphasize that the large angular size of the isolation region 
 for low redshift systems artificially removes them from a catalog. 
Eliminating the isolation criterion includes these nearby groups 
 at the expense of including compact group candidates 
 that are substructures in massive systems. 
However, the compact group candidates within massive systems 
 or in dense regions can be removed after the initial selection. 
In contrast, the low redshift systems cannot be recovered.

\begin{figure}
\centering
\includegraphics[width=8.5cm,angle=0]{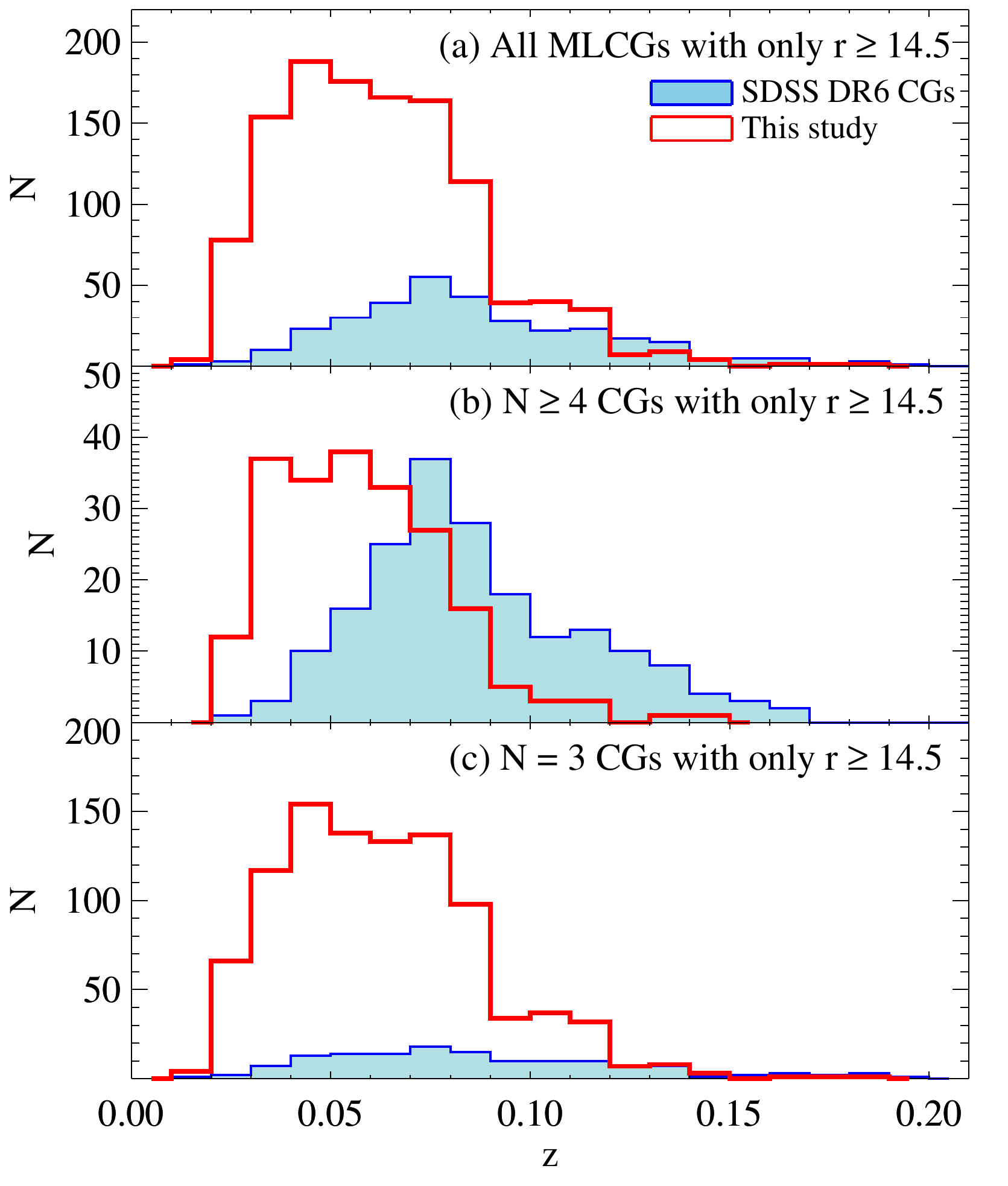}
\caption{
Redshift distribution of the MLCG systems 
 containing only galaxies with $r \geq 14.5$ (open histograms) compared with 
 the SDSS DR6 compact groups (filled histograms, \citealp{Soh15}). }
\label{mlz}
\end{figure}

The probability of rejection as a result of the isolation criterion increases with decreasing redshift. 
The lowest redshift compact group candidates have larger angular size
 and a correspondingly more extensive isolation region
 than more distant compact group candidates. 
Figure \ref{comp} (a) shows the difference between 
 the redshift distribution of the MLCG systems and the SDSS DR6 compact groups. 
Because \citet{McC09} remove $r < 14.5$ galaxies, 
 they might miss some nearby groups thus accounting for the difference. 
To test this conjecture,
 Figure \ref{mlz} plots the redshift distribution of the MLCG systems
 containing only $r \geq 14.5$ galaxies in analogy with the SDSS DR6 sample.
Although we remove 407 MLCG systems containing bright members, 
 we still identify plenty of compact groups in the local universe ($z < 0.05$)
 where SDSS DR6 compact systems are absent or rare. 
This comparison clearly demonstrates that
 the isolation criterion selectively removes low redshift compact group
 candidates from the catalog. 
    
Because we do not apply an isolation criterion in the selection algorithm, 
 we include compact group candidates in dense surroundings (see Figure \ref{nc}). 
Indeed, the fraction of the MLCG systems in dense environments ($N_{C} > 7$) 
 is larger than for the SDSS DR6 compact groups at all redshifts. 
Interlopers may occur more often in high-density regions 
 thus producing some of the MLCG systems with inflated line-of-sight velocity dispersions.  
\citet{Bar07} emphasize that 
 a study of some aspects of galaxy evolution in 
 tight pairs and compact groups 
 requires specialization to relatively low density environments.  
As in pair studies, 
 candidate systems in dense surroundings can be removed after the general group selection.

\subsection{Environmental Effects}

The diverse local environments of compact groups 
 have been discussed previously
 (e.g. \citealp{Ram94, Rib98, And05, Men11, Pom12, Dia12, Dia15}). 
The fraction of embedded compact groups changes significantly 
 depending on the group identification method and on the definition of local environment.
For example,
 \citet{Ram94} suggested that 76\% of 29 Hickson compact groups are embedded, and 
 \citet{Men11} showed that 50\% of the SDSS DR6 compact groups are  
 within $1~h_{70}^{-1}$ Mpc of rich groups. 
In contrast, \citet{Pom12} found that 33\% of the compact groups
 in the second Digitized Palomar Observatory Sky Survey (DPOSS II) 
 are embedded in rich groups.
\citet{Dia15} also found that only 27\% of compact groups
 reside in loose groups in the 2MASS compact group catalog. 
A direct comparison with the MLCG systems is difficult because we use a different environment measure. 

\begin{figure}
\centering
\includegraphics[width=8.5cm,angle=0]{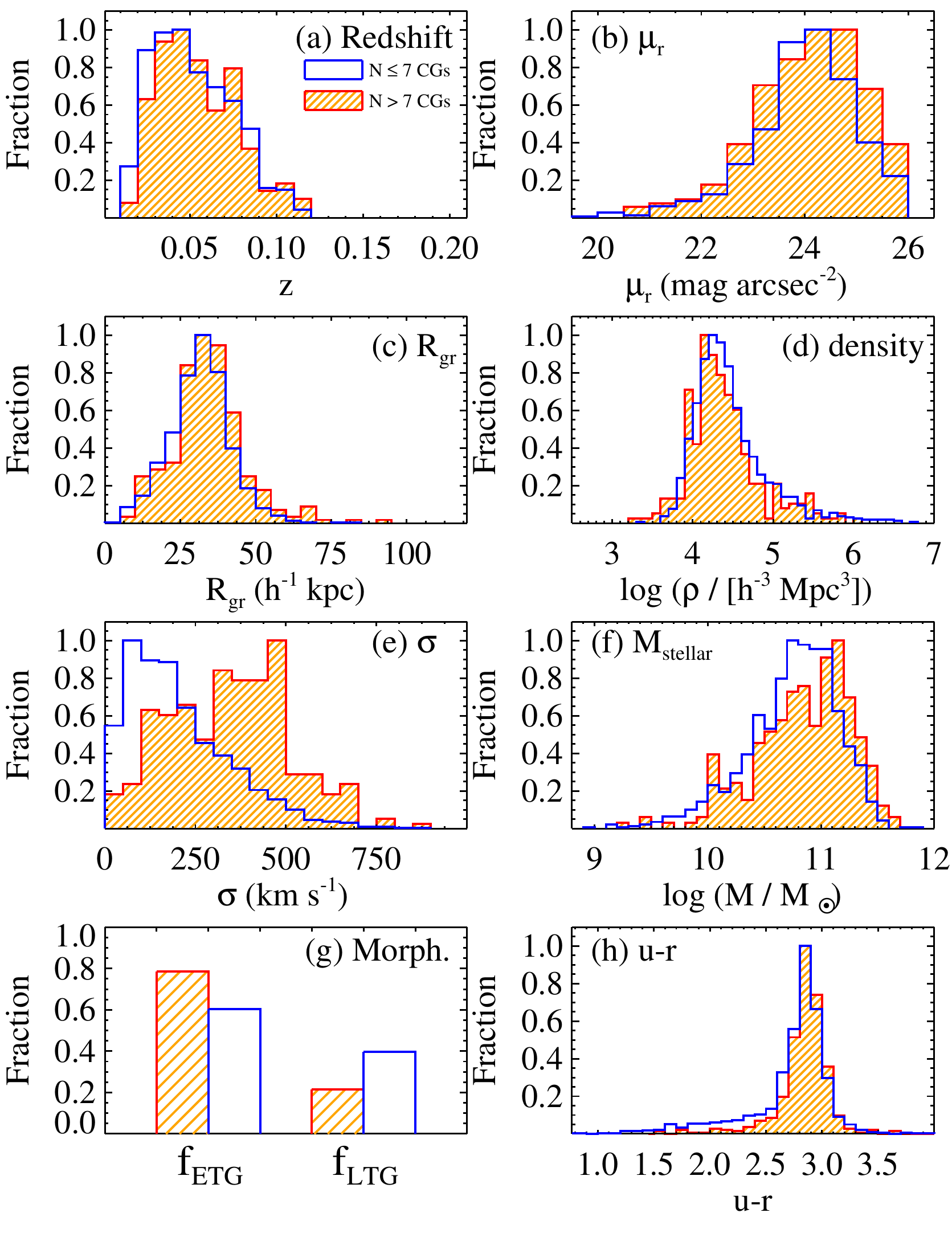}
\caption{
Properties of compact groups as a function of environment ($N_{C}$):  
 (a) redshift, (b) surface brightness, 
 (c) size, (d) galaxy number density, (e) velocity dispersion, (f) stellar mass for
 $N_{C} \leq 7$  (open histogram) and 
 $N_{C} > 7$ (hatched histogram). 
The lower two panels show 
 (g) the fraction of early- and late-type galaxies and 
 (h) the $u-r$ color distribution for $N_{C} \leq 7$ and $N_{C} > 7$ compact groups. }
\label{emb}
\end{figure}

Compact group properties may depend on the local environments. 
We compare the physical properties of groups segregated by $N_{C}$ in Figure \ref{emb}.
Panels (a)-(f) show 
 redshift, $r-$band surface brightness, size, galaxy number density, 
 velocity dispersion, and stellar mass of the MLCG systems. 
The plots show that 
 properties related to the group identification,
 including redshift, surface brightness, size and number density,
 are consistent irrespective of the local environment. 
In contrast with the 2MASS compact group sample \citep{Dia15}, 
 we find no dependence of the MLCG projected size on environment. 
The dependence found by \citet{Dia15} may result from 
 the group identification algorithm (see Section \ref{iso}).
On the other hand, 
 the velocity dispersion and stellar mass of $N_{C} > 7$ MLCG systems
 tend to be larger than for $N_{C} \leq 7$ groups. 
This result is consistent with the comparison between
 `isolated' and `embedded' compact groups
 in the DPOSS II \citep{Pom12} and the SDSS DR6 samples \citep{Soh15}. 
These results are understandable 
 because the interloper fraction may be enhanced in dense environments and 
 because galaxies with greater stellar mass tend to inhabit denser regions 
 (e.g. \citealp{Bol10, Dam15}). 
Table \ref{cgprop_emb} lists 
 the range and the median of the properties of the MLCG systems 
 in different environments. 
 
%%=================================
%%%%   Table 5
%%=================================
\begin{table}
 \begin{center}
  \label{cgprop_emb}
  \caption{Comparison between Compact Groups in Different Environments}
  \begin{tabular}{ccccc}
  \hline
  \multicolumn{2}{c}{Properties} & All CGs & $N_{C} \leq 7$ CGs & $N_{C} > 7$ CGs \\
  \hline 
  \multirow{2}{*}{z}            & Range  & [0.011, 0.188]    & [0.014, 0.112]    & [0.019, 0.112] \\
                                & median & $0.050 \pm 0.001$ & $0.049 \pm 0.001$ & $0.053 \pm 0.003$ \\
  \hline
  $R_{gr}$			            & Range  & [4.9,  94.1]      & [4.9,  83.7]      & [8.9,  94.1] \\
  ($h^{-1}$ kpc)                & median & $32.0 \pm  0.3$   & $31.8 \pm  0.3$   & $33.6 \pm  0.8$ \\
  \hline
  $\log \rho$ 			        & Range  & [3.29, 6.77]      & [3.39, 6.77]      & [3.29, 6.00] \\
  ($h^{3} {\rm Mpc}^{-3}$)      & median & $4.36 \pm 0.01$   & $4.37 \pm 0.01$   & $4.31 \pm 0.03$ \\
  \hline
  $\sigma$                      & Range  & [1,  879]         & [1,  866]         & [1,  879] \\
  (km s$^{-1}$)                 & median & $194 \pm  4$      & $171 \pm  4$      & $353 \pm 14$ \\
  \hline
  \multicolumn{2}{c}{$f_{ETG} (\%)$}     & $64.0 \pm 0.7$    & $60.4 \pm 0.8$    & $78.6 \pm 1.3$ \\
  \hline  
  \end{tabular}
 \end{center}
\end{table}

We also compare the properties of member galaxies
 in the $N_{C} \leq 7$ and $N_{C} > 7$ MLCG systems
 in Figure \ref{emb} (g) and (h). 
The properties of member galaxies including 
 the fraction of early-type galaxies and the $u-r$ color distribution
 differ.
The fraction of early-type galaxies is higher in 
 $N_{C} > 7$ compact groups ($78.7 \pm 1.4 \%$) than in 
 $N_{C} \leq 7$ compact groups ($60.4 \pm 0.7 \%$). 
The member galaxies in $N_{C} > 7$ compact groups are, on average, 
 redder than those in $N_{C} \leq 7$ compact groups 
 as one might expect 
 based on the known relations between galaxy properties and local density (e.g. \citealp{Par07, Bla09}).

The differences in physical properties 
 are qualitatively insensitive to the definition of `dense' environments
 for $N_{C} = 3, 5, 7, 10, 15, 20$. 
The MLCG systems in denser environments show
 larger velocity dispersion, larger stellar mass
 and higher early-type fraction; 
 other properties are essentially environment independent. 

\section{VOLUME LIMITED SAMPLES} \label{vlsam} 

Selection of compact group candidates from a complete redshift survey 
 offers a unique opportunity for construction of volume-limited subcatalogs. 
If the underlying galaxy catalog were complete, 
 the number density of compact groups would be a robust estimate of their true physical space density.
Compact group candidates in the volume-limited catalogs V1CG and V2CG (See Section \ref{vol}) 
 should have properties 
 that are essentially redshift independent. 
In contrast with the MLCG systems, 
 comparison of the total stellar masses for the V1CG and V2CG catalogs require 
 negligible relative correction for the unobserved portion of the mass function.  

\begin{figure}
\centering
\includegraphics[width=8.5cm,angle=0]{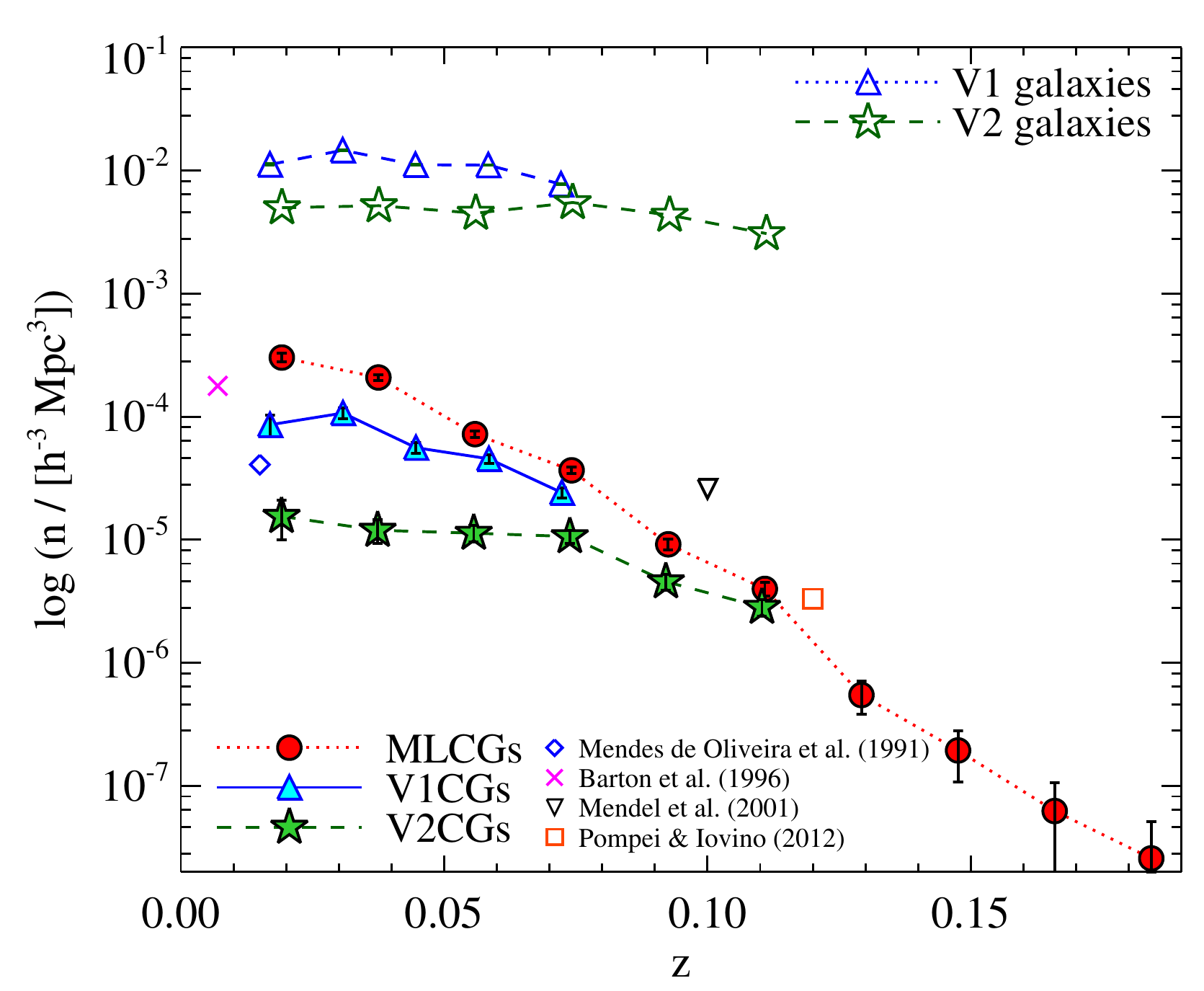}
\caption{
Abundance of the MLCG systems (filled circles), V1CG systems (filled triangles), and V2CG systems (filled stars)
 as a function of redshift.
We also plot the abundance of V1 galaxies (open triangles) and V2 galaxies (open stars); 
 these number densities are nearly constant as expected. 
The number densities of the compact group samples decline artificially with redshift 
 because of SDSS fiber-positioning constraints. }
\label{abun}
\end{figure}

The observed number density of galaxies in the underlying volume-limited catalogs
 is roughly constant (Figure \ref{abun}). 
However the number density for both V1CG and V2CG systems declines with redshift. 
This decline cannot be a result of evolution: unless compact groups are replenished, 
 their number density must decline with cosmic time \citep{Dia94}. 
For the V1CG systems the decline in Figure \ref{abun} begins at $z \sim 0.04$; 
 for the V2CG systems it begins at $z \sim 0.07$.  
This behavior reflects the impact of the SDSS fiber-positioning constraints. 
\citet{She16} showed that 
 the fraction of missing close pairs is 
 a function of both redshift and apparent magnitude. 
The number density of the V2CG systems declines significantly at a redshift 
 where the fiber exclusion radius of 55 arcsec becomes comparable with 
 the projected linking length of $50~h^{-1}$ kpc we apply. 
This behavior is similar to the behavior in Figure 2 of \citet{She16}. 
The steeper decline for the V1CG systems probably reflects 
 the selection against the lower luminosity objects in these groups. 

As a result of the fiber-positioning constraints, 
 we cannot be confident that the compact group catalog is complete at any redshift. 
However it is interesting to note that
 the number density at the lowest redshifts is consistent with 
 previous determinations \citep{MdO91, Bar96, Soh15}. 
Figure \ref{abun} also shows number density estimates for 
 \citet{Men11} and \citet{Pom12} catalogs.
These abundances appear to track the artificially declining abundance of the MLCG. 
This consistency suggests that the previous catalogs may also be incomplete.

\begin{figure}
\centering
\includegraphics[width=8.5cm,angle=0]{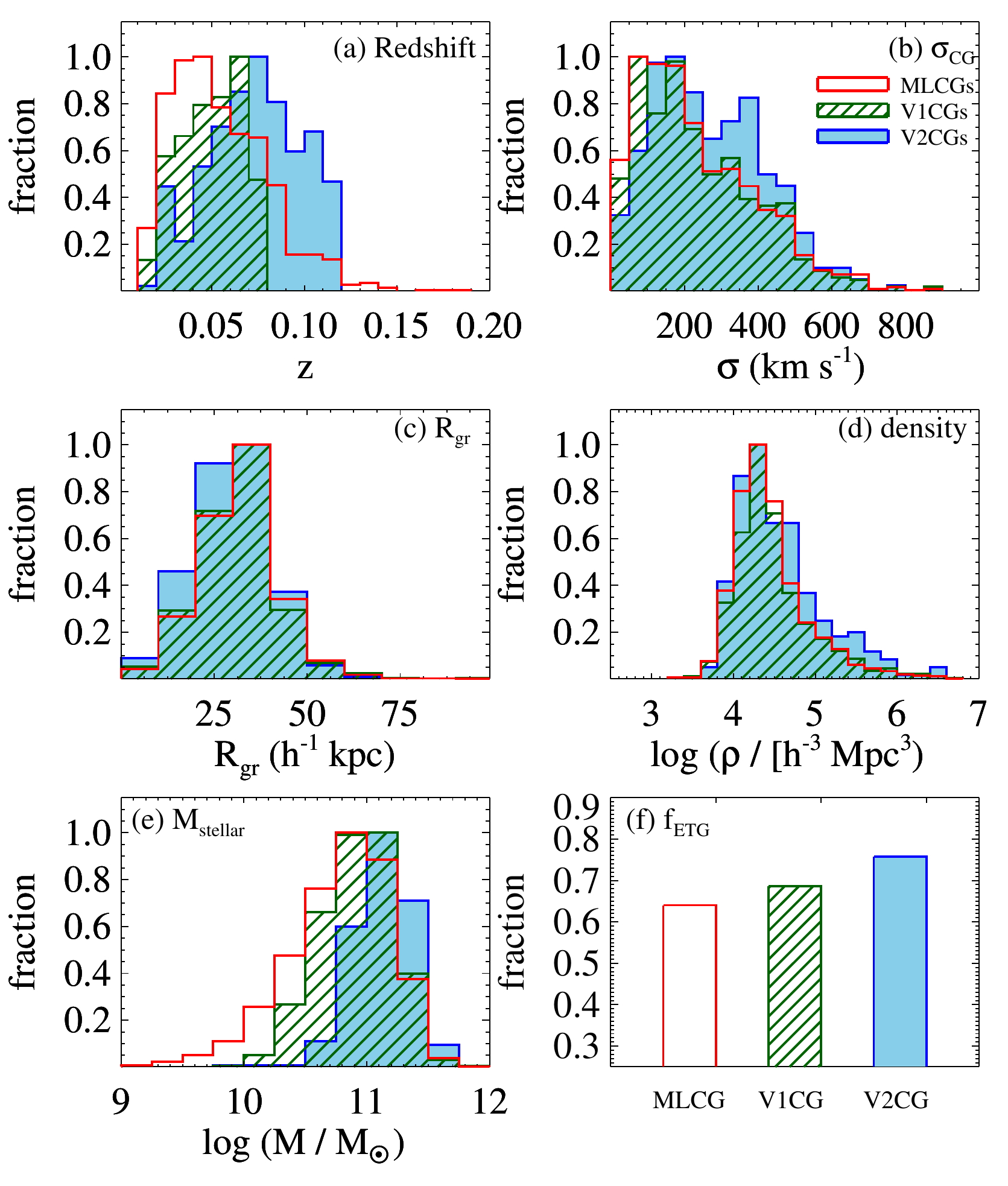}
\caption{
Comparison of physical properties of the MLCG systems (open histograms), 
 V1CGs (hatched histograms), and V2CGs (filled histograms) 
 including (a) mean redshift, (b) line-of-sight velocity dispersion ($\sigma_{CG}$), (c) physical radius, 
 (d) galaxy number density, (e) total stellar mass and (g) fraction of early type galaxies. }
\label{volcomp}
\end{figure}

Although the volume-limited samples are not complete, 
 they still provide a set of systems with a similar range of physical properties 
 throughout the sample redshift range. 
Figure \ref{volcomp} shows normalized histograms of the 
 distributions of properties for the MLCG, V1CG, and V2CG catalogs. 
The difference in redshift distribution simply reflects the selection. 
The distribution of velocity dispersions for the V2CGs appears double peaked. 
The groups in the higher velocity dispersion peak are typically in denser regions with $N_{C} \sim 5$.
As expected based on the selection, the V2CG systems have larger total stellar masses than the V1CG systems. 
The low stellar mass tail present among the MLCG systems is absent from the volume-limited samples. 

\begin{figure*}
\centering
\includegraphics[scale=0.9]{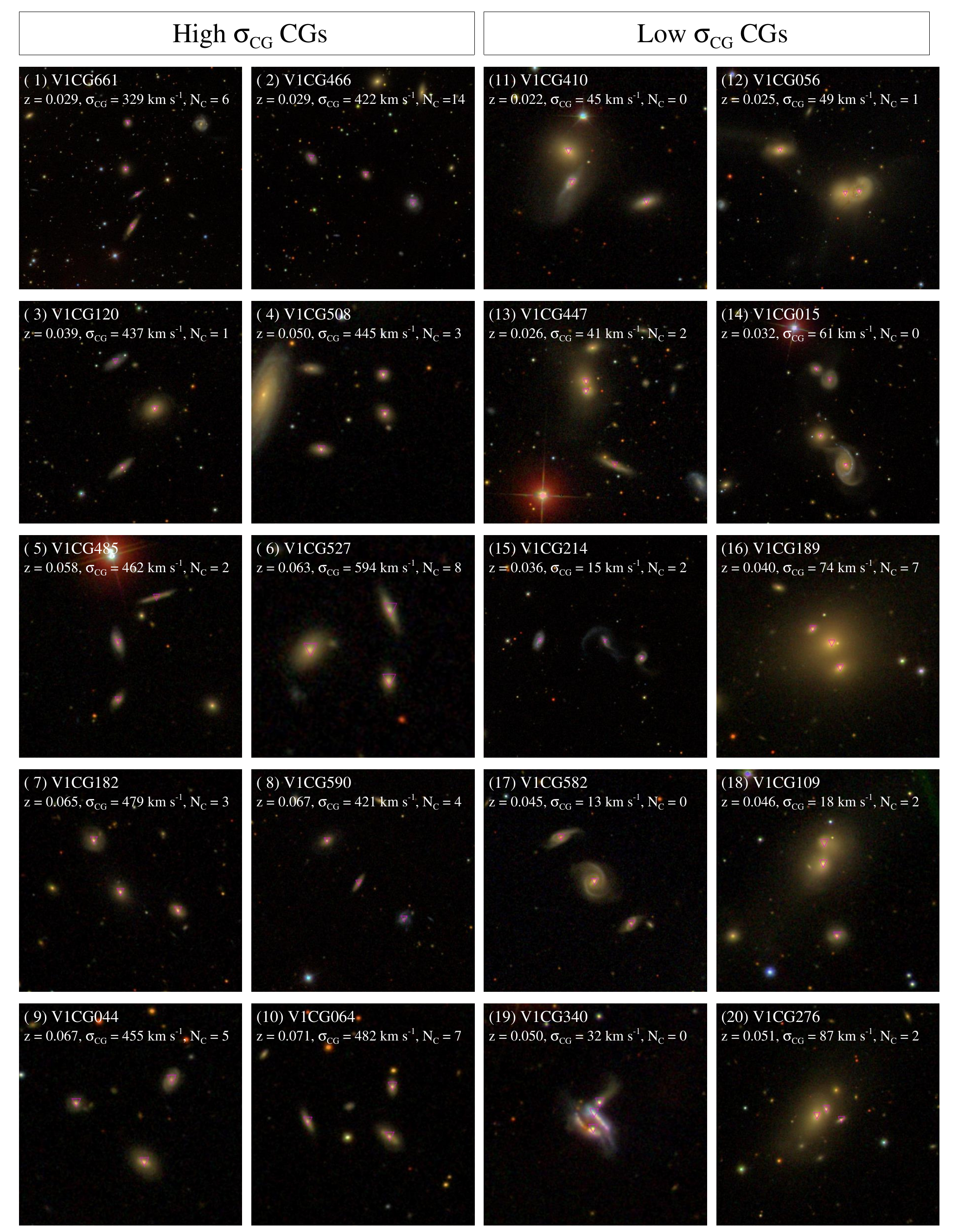}
\caption{Sample images of V1CGs. 
The left two columns shows examples with 
 $\sigma_{CG} \geq 300 \kms$. 
The right two columns show images of groups with 
 $\sigma_{CG} < 100 \kms$. }
\label{thumb}
\end{figure*}

Figure \ref{thumb} provides 
 a qualitative guide to some of the interesting properties
 of the candidate systems in the catalog. 
Systems with low line-of-sight velocity dispersion ($\sigma_{CG} < 100 \kms$) 
 are {\it a prior} likely to be bound. 
Remarkably the montage in Figure \ref{thumb} shows that nearly all of these systems, 
 even with the shallow SDSS photometry, 
 show signs of dynamical interaction. 
Late-type galaxies often have tidal tails (e.g. panel 11, 14, 15, 19);
 early type galaxies are often apparently embedded in a common halo (e.g. panel 12, 13, 16, 18, 20). 
Each panel of the montage lists $N_{C}$. 
Most of these objects have low $N_{C}$ suggesting that
 they are excellent candidates for the study of relatively isolated, 
 probably interacting/merging systems. 
V1CG189 is the main exception with an $N_{C} = 7$; 
 it is probably a substructure in a richer system.

In contrast systems with high velocity dispersion ($\geq 300\kms$) 
 show little or no evidence of obvious dynamical interaction.  
Deeper observations might reveal such features, 
 but with the current data there is no way of judging 
 whether the system is a true bound system. 
It is interesting to note that 
 the range of $N_{C}$ for these systems is larger
 than for the massive systems; 
 three of these example systems have $N_{C} \geq 7$. 
In these cases, the velocity dispersion may well be inflated 
 by one or more interlopers from the cluster. 
For velocity dispersions $100 < \sigma_{CG} < 300 \kms$, the candidate systems 
 show a mix of qualitative visual properties along with a mix of environments.  

The volume-limited samples provide a platform for 
 more detailed studies of the physical properties of 
 a homogeneous sample of compact group candidates 
 including spectroscopic properties, dynamical studies (e.g. \citealp{Bar96, Pom12, Soh15}), 
 deeper photometric observations (e.g. \citealp{Bro15}), and 
 observations in other wavebands from the radio to the X-ray (e.g. \citealp{Des14, Wal16}). 
Many of these groups have an angular size comparable to the MANGA field of view 
 ($\sim32"$, \citealp{Bun15}). 
Thus detailed spatially resolved spectroscopy of these systems 
 could provide fresh insight into the apparent dynamical interactions among the members.  

%=============================================================
\section{CONCLUSION}

We apply an FoF method
 to an enhanced SDSS DR12 spectroscopic catalog to 
 construct a catalog of $1588~N\ge3$ compact groups
 containing 5179 member galaxies and covering the redshift range $0.01 < z < 0.19$. 
The approach to the construction of this catalog is similar to \citet{Bar96}. 
However, the new catalog contains 18 times as many systems 
 and reached to 3 times the depth.
These two catalogs are unique in their derivation 
 from dense and nearly complete redshift surveys. 
The general properties of these spectroscopic compact groups 
 including their velocity dispersions, sizes, densities, and 
 galaxy population are similar to those
 previously selected from photometric datasets.

We use a fixed projected physical spatial and rest frame line-of-sight velocity linking lengths
 to generate a catalog where group projected size and density are redshift independent.  
Even with fixed selection parameters, a frequently applied isolation criterion produces 
 an artificial increase in compact group size with redshift in many previous catalogs. 

Application of an isolation criterion also mitigates 
 against the inclusion of nearby groups in a catalog. 
The catalogs we construct contain many more compact group candidates 
 at $z\lesssim 0.05$ than previous catalogs. 
Many of these systems show 
 obvious evidence for current tidal interactions among the member galaxies. 

When we explore the catalog as a whole, 
 the fraction of obviously interacting systems is striking for groups
 with velocity dispersions $\lesssim 100 \kms$.
For group candidates with the largest dispersions ($\gtrsim 300 \kms$),
 there is little evidence for these interactions 
 perhaps suggesting some interloper contamination. 
Deeper photometric observations are necessary to explore this issue further. 

The catalog includes systems in a range of environments. 
Group properties depend somewhat on the environment in the sense 
 that groups in dense environments contain a larger fraction of early-type galaxies as expected.

In contrast, with earlier investigations, 
 we construct volume-limited catalogs 
 in addition to the primary magnitude-limited sample. 
The volume-limited catalogs have several important advantages. 
The systems should have similar physical properties throughout the redshift range. 
Unlike the systems in the magnitude-limited catalog, 
 the volume- limited systems have total stellar masses nearly independent of redshift.  
It is important here that there is
 little correction for the missing end of the mass function 
 because the samples are volume-limited. 

The volume-limited samples are potentially powerful probes of the number density of compact groups. 
At low redshift our number density $\sim 10^{-4} \hmpc$ agrees with \citep{MdO91,Bar96}. 
The volume-limited sample number density as a function of redshift provides 
 an important probe of the impact of the SDSS fiber placement constraint;
 a decline in the number density is apparent at redshifts corresponding to the constraint.
This fiber placement constraint is currently 
 a fundamental limitation on the interpretation of the number density 
 as a function of redshift. 
Eliminating this issue (see \citealp{She16}) could enable insights into 
 the number density evolution of compact groups based on volume-limited samples.

The abundant systems in the volume-limited compact group catalogs at $z \leq 0.07$ offer a rich foundation 
 for detailed investigations of spatially resolved spectroscopic measures 
 of star formation histories and stellar population ages. 
At these redshifts the surface brightness of tidal features is high enough 
 to support extensive exploration of the internal galaxy kinematics. 
Taken together dynamical studies coupled with other age indications 
 may provide a route to solving the mystery of the existence of compact groups.

%%% ACKNOWLEDGMENTS (IF ANY) %%%%%%%%%%%%%%%%%%%%%%%%%%%%%%%%%%%%%%%%
\acknowledgments

This work was supported by the National Research Foundation of Korea (NRF) grant
 funded by the Korea Government (MSIP) (No.2013R1A2A2A05005120).
J.S. was supported by 
 Global Ph.D. Fellowship Program through an NRF funded by the MEST (No. 2011-0007215). 
The research of J.S., M.J.G., and HJZ is supported by the Smithsonian Institution.
H.J.Z. gratefully acknowledges the support of the Clay Fellowship. 

Funding for SDSS-III has been provided by the Alfred P. Sloan Foundation, 
 the Participating Institutions, the National Science Foundation, 
 and the U.S. Department of Energy Office of Science. 
The SDSS-III web site is http://www.sdss3.org/.
SDSS-III is managed by the Astrophysical Research Consortium 
 for the Participating Institutions of the SDSS-III Collaboration 
 including the University of Arizona, the Brazilian Participation Group, 
 Brookhaven National Laboratory, Carnegie Mellon University, University of Florida, 
 the French Participation Group, the German Participation Group, Harvard University, 
 the Instituto de Astrofisica de Canarias, the Michigan State/Notre Dame/JINA Participation Group, 
 Johns Hopkins University, Lawrence Berkeley National Laboratory, 
 Max Planck Institute for Astrophysics, Max Planck Institute for Extraterrestrial Physics, 
 New Mexico State University, New York University, Ohio State University, 
 Pennsylvania State University, University of Portsmouth, Princeton University, 
 the Spanish Participation Group, University of Tokyo, University of Utah, 
 Vanderbilt University, University of Virginia, University of Washington, and Yale University.

\appendix
\section{\label{app}COMPACT GROUP CANDIDATES ELIMINATED BY POPULATION AND SURFACE BRIGHTNESS SELECTION}
%%\clearpage

In the construction of the MLCG catalog 
 we do not apply Hickson's original isolation criterion (See Sections \ref{fof}). 
We do, however apply 
 the `population' criterion ($N~(\Delta r \leq 3) \geq 3$). 
This criterion
 removes 321 systems that include a bright galaxy surrounded by 
 much fainter apparent satellite galaxies at very similar redshift.  
We also apply Hickson's `compactness' criterion 
 ($\mu_{r} \leq 26 {\rm~mag~arcsec}^{-2})$ which excludes a further 120 groups. 
This criterion excludes systems containing galaxies of 
 much lower surface brightness than 
 the typical systems over the redshift range covered by the MLCG. 
Here we briefly examine subsets of the systems removed by these two criteria. 
These systems may be useful for applications other than 
 the study of traditional compact group candidates.
 
\begin{figure*}
\centering
\includegraphics[scale=0.7]{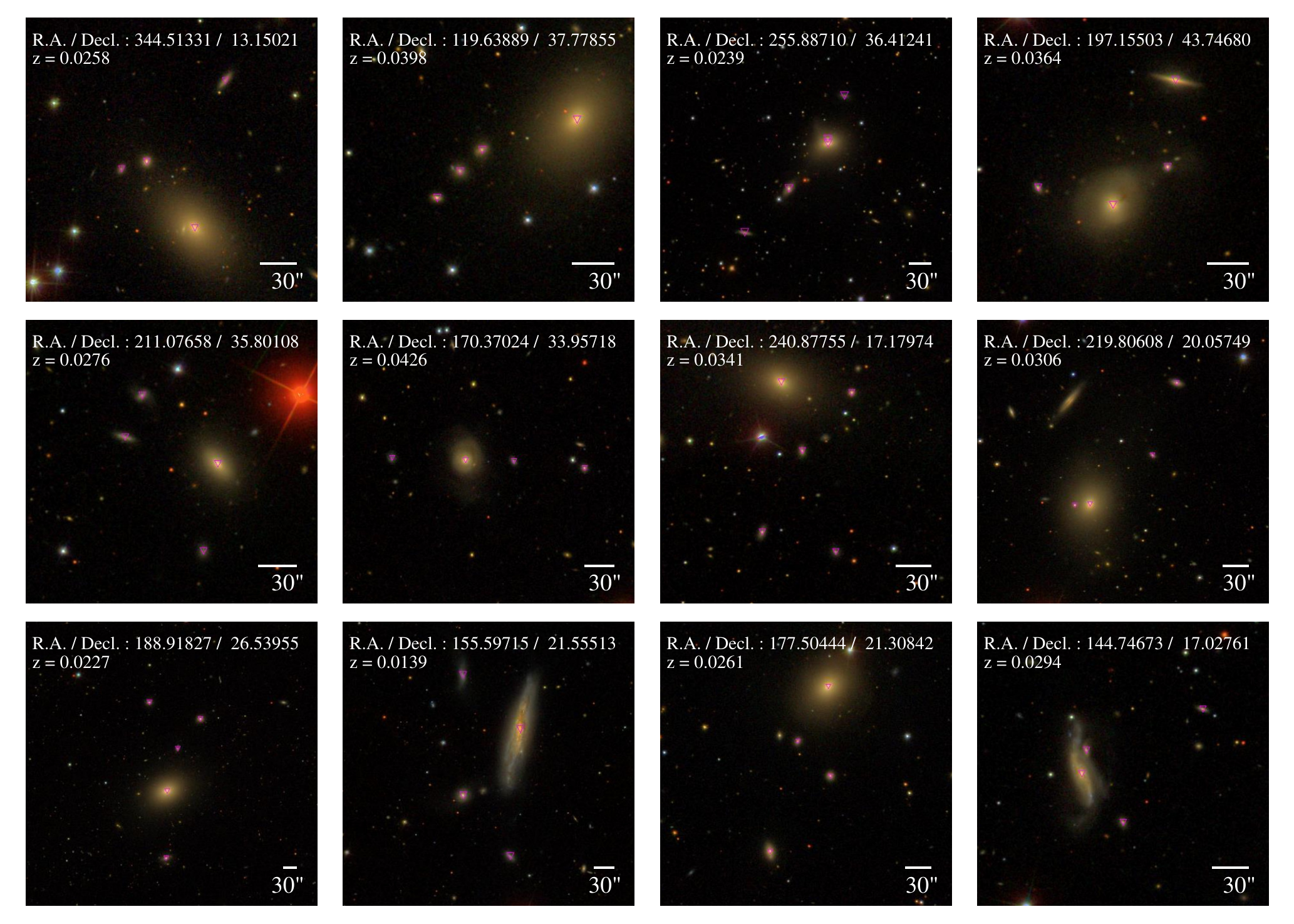}
\caption{
Sample images of compact group candidates rejected by the population criterion.
They contain member galaxies differing from the brightest member by $\Delta r > 3$.
 Magenta symbols mark the member galaxies. 
 The white bar shows a 30 arcsec scale at the mean group redshift. }
\label{dominant}
\end{figure*}

Figure \ref{dominant} shows 
 example images of some the systems that 
 violate the `population' criterion. 
As these images show, 
 these objects have one dominant bright galaxy
 surrounded by much fainter satellites. 
Unlike the more standard compact group candidates we discuss in the body of this paper, 
 these group candidates are not comprised of galaxies 
 with comparable stellar masses.
 
Systems like those in Figure \ref{dominant} have been 
 the basis for investigation 
 the dark matter halo of the dominant galaxy \citep{Zar93, Nor08, Woj13}. 
To measure the distribution of 
 dark matter in the dominant galaxy halo, 
 \citet{Zar93} compiled the redshifts for the satellites of 45 apparently isolated spiral galaxies; 
 the typical systems they investigated have one to four satellites. 
Other investigators have since used 
 larger samples satellites to revisit the constraints 
 on the mass distribution in galaxy dark matter halos 
 \citep{Hol69, Sal05, Pre11, Wan14}.
The FoF we apply to the SDSS DR12 yields 321 similar systems 
 that could be used for further investigation of these issues. 
We include the catalog in Table 8 and Table 9.% \ref{dom_gr} and Table \ref{dom_gal}.

\begin{figure*}
\centering
\includegraphics[scale=0.7]{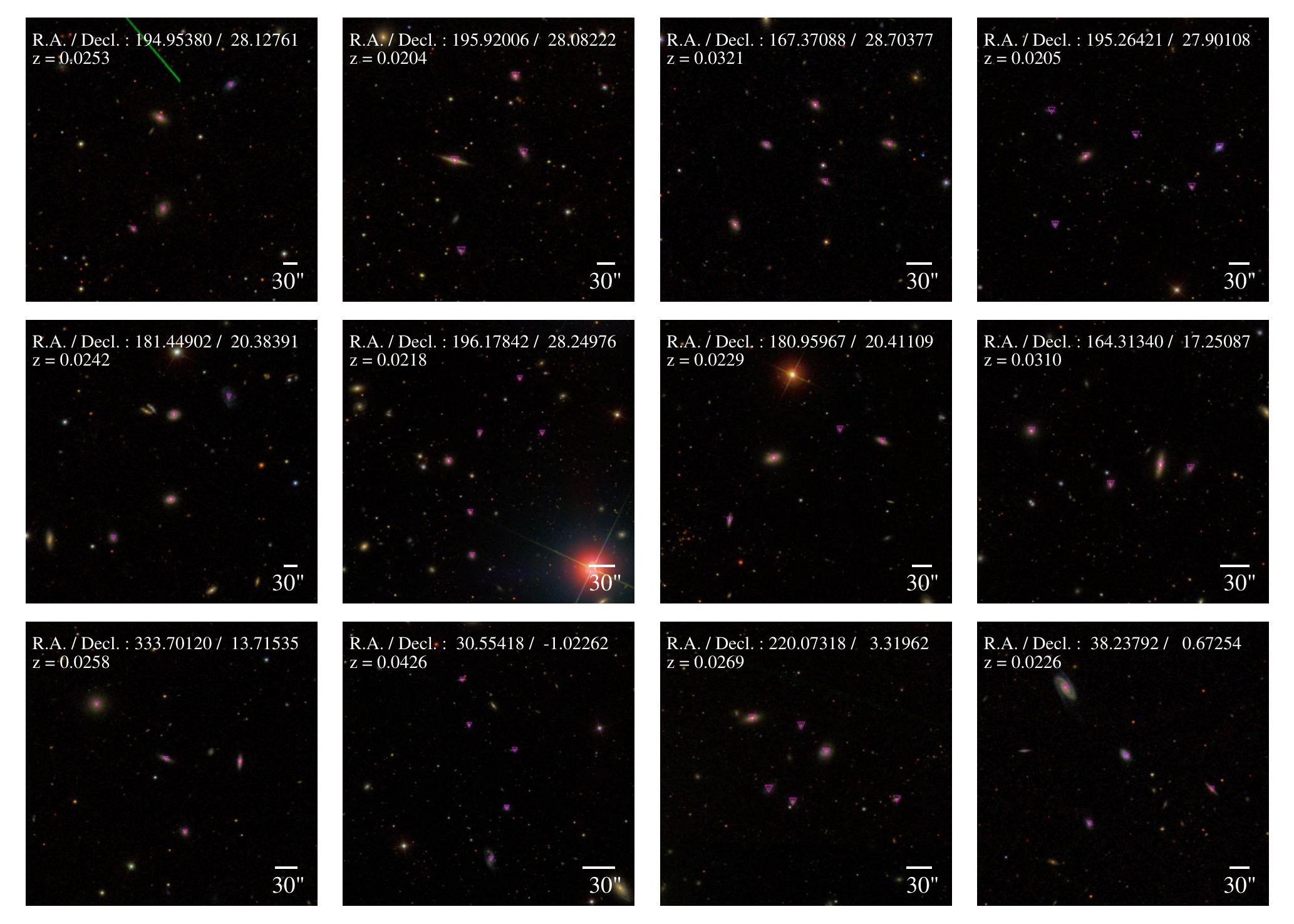}
\caption{Sample images of compact group candidates rejected by the surface brightness criterion:
 $\mu_{r} > 26 {\rm~mag~arcsec^{-2}}$.
 The magenta symbol and the white bar have the same meaning as in as Figure \ref{dominant}. } 
\label{faint}
\end{figure*}

Figure \ref{faint} displays examples of the 120 systems 
 eliminated from compact group catalog
 because of their low surface brightness. 
These systems are comprised of small, low luminosity galaxies. 
These candidate systems show no obvious sign of recent interaction/merger 
 in the SDSS images. 
The search radius we use is essentially tuned to find
 dense systems of galaxies with stellar masses $M_{stellar} \sim 6 \times 10^{9}~{\rm M_{\odot}}$.
Finding candidate interacting systems of low luminosity objects probably requires tighter selection criteria. 
These systems are rare in our catalog because they are only visible in the lowest redshift portion of the catalog; 
 their maximum redshift is $z = 0.05$ and most of them have redshifts around $z \sim 0.03$. 
For completeness we include these objects in Table 10 and Table 11. %\ref{faint_gr} and Table \ref{faint_gal}. 

%%%=================================
%%%%%   Appendix - table 8
%%%=================================
\begin{table*}
 \begin{center}
  \label{dom_gr}
  \caption{Catalog of System Containing a Dominant Galaxy Surrounded by Satellites}
  \begin{tabular}{cccc}
  \hline
  R.A. & Decl. & $N_{\rm galaxy}$ & redshift \\
  \hline
  198.227158 &   1.012775 &  3 & 0.0723  \\
  141.442184 &  11.437011 &  3 & 0.0113  \\
  127.134727 &  30.435843 &  4 & 0.0506  \\
  188.723557 &  47.756451 &  3 & 0.0306  \\
  187.538025 &  47.021252 &  3 & 0.0394  \\
  \hline
  \end{tabular}
 \end{center}
 \tablecomments{The full table is available in the online journal. A portion is shown here for guidance regarding its form and content.} 
\end{table*}
   
%%%=================================
%%%%%   Appendix - table 9
%%%=================================
\begin{table*}
 \begin{center}
  \label{dom_gal}
  \caption{Catalog of Galaxies in the System Containing a Dominant Galaxy Surrounded by Satellites}
  \begin{tabular}{lcccc}
  \hline
  Object ID & R.A. & Decl. & $r$ & redshift \\
  \hline
  1237648705657307347 & 198.229294 &   1.010990 & 17.42 & $0.0727 \pm 0.00002$ \\
  1237648705657307315 & 198.218872 &   1.019821 & 16.48 & $0.0704 \pm 0.00002$ \\
  1237661070318370904 & 141.447220 &  11.424596 & 12.45 & $0.0125 \pm 0.00001$ \\
  1237661070318370911 & 141.452072 &  11.454233 & 15.11 & $0.0108 \pm 0.00001$ \\
  1237661070318370902 & 141.427277 &  11.432204 & 11.76 & $0.0107 \pm 0.00004$ \\ 
  \hline
  \end{tabular}
 \end{center}
 \tablecomments{The full table is available in the online journal. A portion is shown here for guidance regarding its form and content.}
\end{table*}
  
%%%=================================
%%%%%   Appendix - table 10
%%%=================================
\begin{table*}
 \begin{center}
  \label{faint_gr}
  \caption{Catalog of System with Low Surface Brightness ($\mu_{r} > 26 {\rm~mag~arcsec^{-2}}$)}
  \begin{tabular}{cccc}
  \hline  
  R.A. & Decl. & $N_{\rm galaxy}$ & redshift \\
  \hline
  140.011292 &  33.666660 &  5 & 0.0222  \\
  141.283279 &  11.557135 &  3 & 0.0121  \\
  247.183990 &  39.646721 &  7 & 0.0292  \\
  149.441193 &  36.060726 &  3 & 0.0267  \\
  247.305191 &  39.599922 &  4 & 0.0327  \\
  \hline
  \end{tabular}
 \end{center}
 \tablecomments{The full table is available in the online journal. A portion is shown here for guidance regarding its form and content.}
\end{table*}

%%%=================================
%%%%%   Appendix - table 11
%%%=================================
\begin{table*}
 \begin{center}
  \label{faint_gal}
  \caption{Catalog of Galaxies in Systems with Low Surface Brightness ($\mu_{r} > 26 {\rm~mag~arcsec^{-2}}$) } 
  \begin{tabular}{lcccc}
  \hline
  Object ID & R.A. & Decl. & $r$ & redshift \\
  \hline
  1237661383844036884 & 140.035767 &  33.661671 & 15.27 & $0.0207 \pm 0.00001$ \\
  1237661383844036895 & 140.053909 &  33.631790 & 16.84 & $0.0221 \pm 0.00001$ \\
  1237661126155436283 & 139.988480 &  33.679878 & 17.70 & $0.0226 \pm 0.00002$ \\
  1237661126155501585 & 140.010468 &  33.710468 & 14.28 & $0.0233 \pm 0.00001$ \\
  1237664870286098561 & 139.967880 &  33.649494 & 15.13 & $0.0222 \pm 0.00001$ \\  
  \hline
  \end{tabular}
 \end{center}
 \tablecomments{
The full table is available in the online journal. A portion is shown here for guidance regarding its form and content.}
\end{table*}

%=============================================================================================================
%  Bibliograph - References
%=============================================================================================================

\end{document}